\documentclass[aip,jcp,preprint,nofootinbib,endfloats]{revtex4}
\usepackage{epsfig,epsf,subfigure,amsmath,amssymb}
\usepackage{color,ulem,hyperref}
\usepackage{overpic}
\usepackage{graphicx}
\usepackage{dcolumn}
\usepackage{bm}
\usepackage{pslatex}

\begin{document}
\author{Santosh Mogurampelly}\thanks{To whom correspondence should be addressed}
\email{santosh@physics.iisc.ernet.in}
\author{Prabal K.~Maiti}
\email{maiti@physics.iisc.ernet.in}
\affiliation{
Centre for Condensed Matter Theory, Department of Physics, Indian Institute of Science,
Bangalore 560012, India}
\title{Translocation and encapsulation of siRNA inside carbon nanotubes}


\begin{abstract}
We report spontaneous translocation
of small interfering RNA (siRNA) inside 
carbon nanotubes (CNTs) of various diameters 
and chirality using all atom molecular 
dynamics (MD) simulations with
explicit solvent. We use Umbrella sampling 
method to calculate the free energy landscape 
of the siRNA entry and translocation
event. Free energy profiles shows that 
siRNA gains free energy while translocating 
inside CNT and barrier for siRNA exit from 
CNT ranges from 40 to 110 kcal/mol depending 
on CNT chirality and salt concentration.
The translocation time $\tau$ decreases with 
the increase of CNT diameter with a critical 
diameter of 24 \AA~ for the translocation.
In contrast, double strand DNA (dsDNA) of the same 
sequence does not translocate inside CNT
due to large {free energy} barrier for the translocation.
This study helps in understanding the nucleic 
acid transport through nanopores at microscopic 
level and may help designing carbon nanotube
based sensor for siRNA.
\end{abstract} 
\maketitle
\pagebreak
\section{Introduction}
Small interfering RNA (siRNA) is 21-23 nucleotides 
long double stranded RNA (dsRNA) which has been 
demonstrated to treat diseases like cancer and 
hepatitis. For such medical applications, siRNA 
has to be delivered inside the cell without degradation
by complexing it with delivering vectors. 
However, efficient method for the delivery
of siRNA to the target cell is lacking. 
Carbon nanotubes (CNTs) are cylindrical 
rolled graphene sheets
which {have} emerged as new nano material to be
used for biomedical applications due to their 
potential strengths \cite{liu1,liu3,kam,zhang}.
CNTs have ultrahigh surface area          
which make them suitable drug delivery vectors
{through} specific binding of biomolecules, drugs on 
their outer surface. This unique feature along 
with the other physical, thermal and optical
properties enable high {through}put drug delivery
and medical imaging \cite{liu3}.
CNTs are not soluble in many solvents {and also} 
known to be toxic to the cell.
This problem has been addressed by covalent 
or non-covalent surface functionalization of CNTs
\cite{liu1,liu3,kam,zhang}. Delivery of RNA by 
translocating RNA-CNT hybrid through membranes 
of MCF7 breast cancer cells has also been studied \cite{lu2004}.
siRNA delivery through pristine and functionalized
CNT have been investigated by several groups in 
recent years. In this delivery approach siRNA is 
adsorbed on the CNT. After transfecting 
siRNA-CNT hybrid inside cells, the siRNA
dissociates from the CNT. Recently we have shown that
siRNA can adsorb onto the surface of the CNT and graphene 
by unzipping its base-pairs and binds strongly with 
high stability which can be used to deliver siRNA \cite{santoshjcp,santoshgraphene}.
All these studies are concerned with 
siRNA adsorption/binding on CNT surface only.
In this paper, we propose encapsulation of siRNA
inside CNT as a possible delivery mechanisms.\\

Transport of biomolecules across cell membranes 
and nanopores is an important process in living 
organisms \cite{hemant_softmatter}.
Single stranded DNA (ssDNA) transport in biologically
occurred nanometer scale$~\alpha$-hemolysin ion channel has been
observed by Kasianowicz {\it{et. al.}} \cite{kasianowicz1996}
in 1996. They observed that the translocation of ssDNA 
depends on the DNA concentration, applied voltage 
and structure of the nanopore. Since then, biomolecular
transport in nanopores has become a major research
activity both {from} fundamental and application point 
of view \cite{henrickson,lubenskinelson1999,meller2001}.
Polymer translocation through membranes \cite{sung1996} 
and nanopores \cite{muthukumar1999,lubenskinelson1999,
muthukumar2001} has been studied theoretically as a 
diffusion process across free energy barrier that 
arises due the chemical potential gradients. 
It has been found that the translocation time 
is proportional to the polymer length.
Sophisticated nanopore based DNA sequencing techniques 
are now-a-days feasible to determine the 
sequence of DNA by measuring ionic currents
as ssDNA translocates through the nanopore
\cite{ceesdekker2007,branton2008}.
However, microscopic level details are necessary
for the better understanding of the
DNA sequencing at {the single} base-pair level
{which is} the 
ultimate goal of DNA sequencing {technology}.
Translocation of ions, nucleic acids through solid state 
nanopores have been studied by Dekker and co-workers
\cite{ceesdekker2006,ceesdekker2007,ceesdekker2009}.
Translocation of oligonucleotides inside CNT 
have been reported as well \cite{gao2003,fan2005nl,gao2007,
gao2008,lim2008,liuscience2010}.
Recently, translocation of DNA through nanopores
in suspended graphene sheets have also been studied 
experimentally that finds applications in DNA 
sequencing \cite{garaj,schneider,christopher,venkatesan2011}.\\

Based on our MD simulation studies, we propose two different methods of siRNA 
delivery using CNTs and graphene. The first mechanism is based on the
strong adsorption of siRNA on the CNT/graphene surface \cite{santoshjcp,santoshgraphene}
and the second one is by encapsulation of siRNA
through translocation inside the CNT. The translocation of siRNA inside a CNT 
has not been studied earlier. In this paper, we 
report for the first time, the translocation 
studies of siRNA inside CNT. This translocation is driven by the 
favorable van der Waals (vdW) interaction 
between the siRNA and CNT. To understand 
the free energy landscape of siRNA translocation 
inside CNT, we have also calculated the free 
energy profile using Umbrella sampling method \cite{torrie,roux,ums} 
for CNTs of different chirality and at 
different salt concentrations. The free 
energy profile indicates that siRNA gains 
energy inside CNT and faces a large energy 
barrier to escape from the interior of CNT.
Insights into siRNA-CNT interaction and structural 
changes will help in understanding the microscopic picture of the 
siRNA delivery mechanism by CNTs.
The translocation time $\tau$ of siRNA
is decreasing with increasing the diameter of the CNT.
Our simulation results can guide
experimental approach for siRNA translocation inside
CNT for sequencing and drug delivery applications {as well}.
The rest of the of the paper is organized as 
follows: in the section \ref{section_methods}, we give the details of 
the simulation methods, followed by the discussion 
on the results of spontaneous translocation.
In section \ref{section_results}, we {discuss 
the results on} free energy landscapes using MM-GBSA as well 
as Umbrella sampling method. Finally in section \ref{section_conclusions},
we give a summary of the main results and conclude.

\section{Methods}
\label{section_methods}
We have built single walled CNT (SWCNT) of different 
chiralities and diameters of length 
approximately 87 \AA.
The diameters $d$ of the built SWCNTs are
27.01 \AA, 26.74 \AA, 25.40 \AA, 24.06 \AA~ and 22.73 \AA~
for (25, 15), (20, 20), (19, 19), (18, 18) and (17, 17)
nanotubes respectively.
We study several systems for nucleic acid translocation 
inside CNTs: (i) siRNA translocation mechanism 
inside (20, 20) CNT at 0 and 107 mM NaCl concentration 
to investigate the effect of salt concentration (ii) siRNA translocation 
inside (19, 19), (18, 18) and (17, 17) CNTs to study
the effect of diameter (iii) siRNA translocation mechanism 
inside (25, 15) to study the effect of chirality
and finally (v) dsDNA (of identical sequence 
except adenine instead of uracil of siRNA) 
translocation mechanism inside (20, 20) CNT
to understand the difference between the interaction 
of siRNA and dsDNA with CNT.
For all these cases, the siRNA/dsDNA was placed 
close to one of the open ends of CNT such that the CNT 
axis and the siRNA axis are parallel. We have used 
the same siRNA sequence and structure as 
described in our earlier publication \cite{santoshjcp,santoshgraphene}.
We have used ff99 force field \cite{duan} for
all the simulations {reported in this paper}.
The siRNA-CNT system is solvated with water and
counterions. To make a charge neutral system
for different salt concentration, appropriate
number of Na$^+$ and Cl$^-$ ions were added. 
Th{e system was} then subjected to standard simulation 
protocol \cite{maiti2004,maiti2006bj} with 
periodic boundary conditions applied in all three directions.
Non-bonded interactions are truncated at 9 \AA~ and
the long range electrostatic interactions were calculated 
with the Particle Mesh Ewald (PME) method \cite{darden}.
Finally for trajectory analysis, simulations are 
performed in constant volume-constant temperature (NVT) ensemble 
up to 100 ns depending on the diameter of CNT.
For the Umbrella sampling method, 120-130 ns of
NVT simulations are performed for each system.
Full system details such as number of NaCl ions, 
water residues, box dimensions and total number of 
atoms are summarized in Table \ref{table1}.
To check the effect of force field, we have 
also performed the simulation with (20, 20)
CNT and siRNA using ff10 force field \cite{yildirim}.

\section{Results and Discussion}
\label{section_results}
\subsection{Translocation of siRNA inside CNTs}
\begin{figure*}
        \centering
        \subfigure[System after equilibration with water and counterions.]
        {
        \includegraphics[height=26mm]{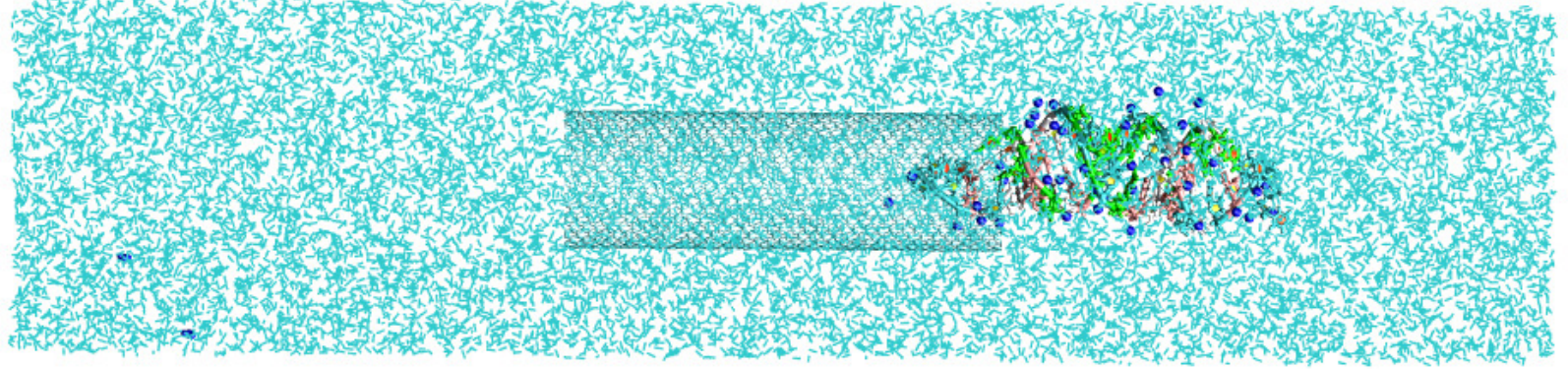}
        \label{sirna-ini}
        }
        \subfigure[3 ns]
        {
        \includegraphics[height=20mm]{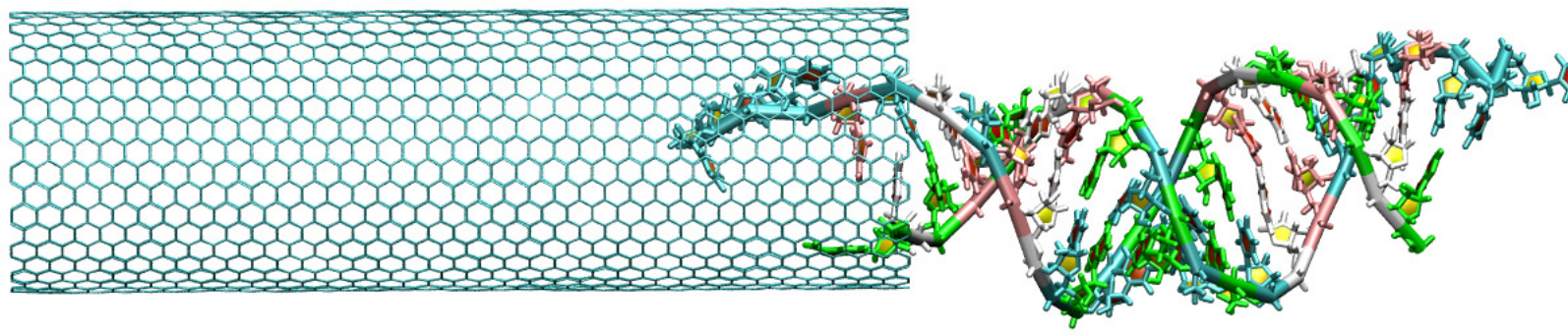}
        \label{3ns_h}
        }
        \subfigure[]
        {
        \includegraphics[height=20mm]{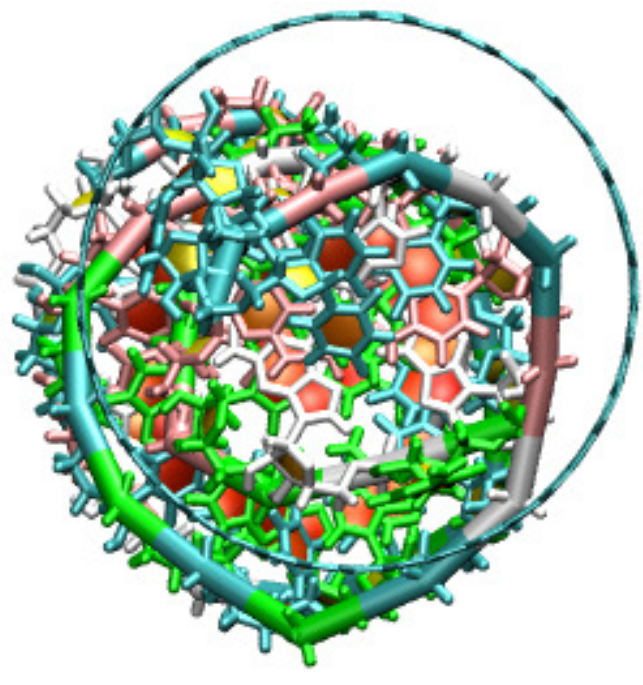}
        \label{3ns_v}
        }\\
        \subfigure[10 ns]
        {
        \includegraphics[height=20mm]{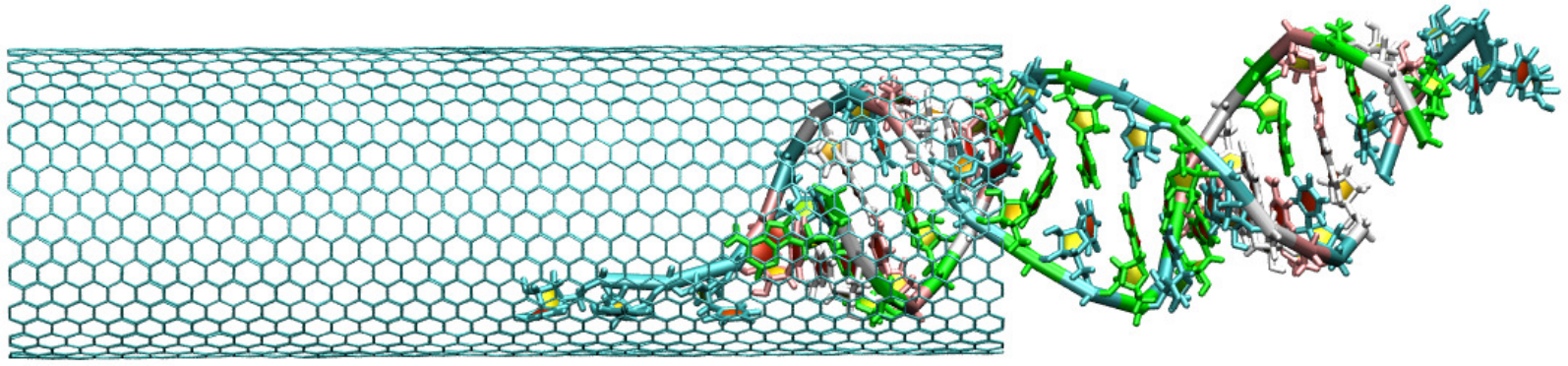}
        \label{10ns_h}
        }
        \subfigure[]
        {
        \includegraphics[height=20mm]{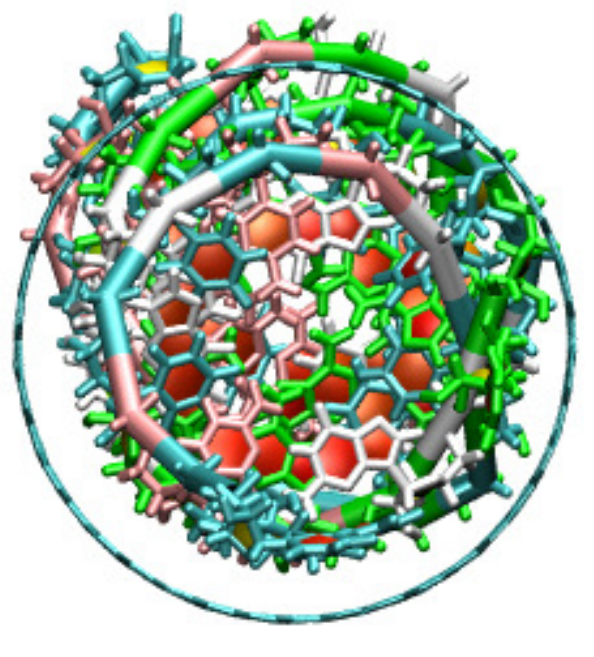}
        \label{10ns_v}
        }\\
        \subfigure[20 ns]
        {
        \includegraphics[height=20mm]{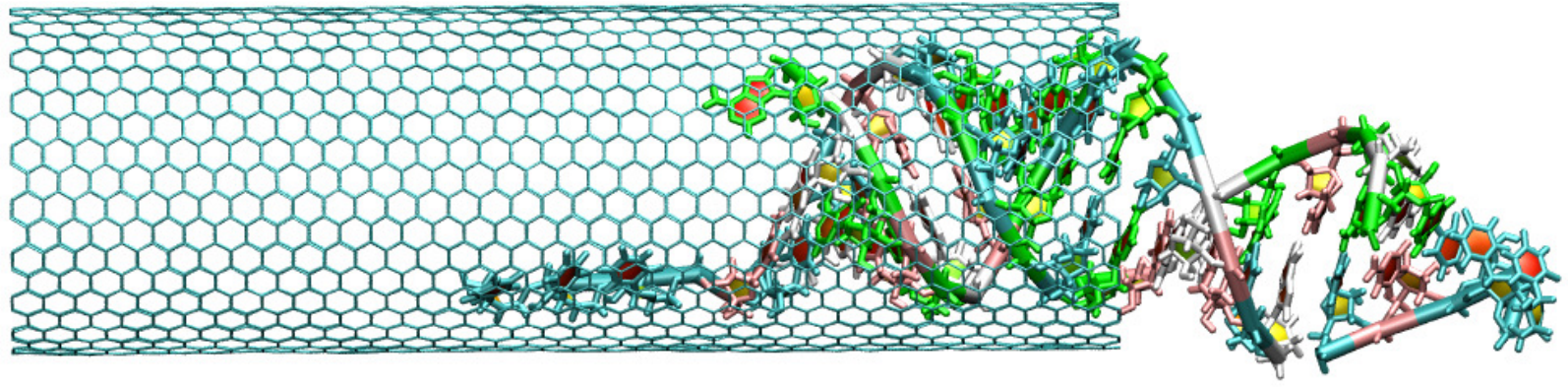}
        \label{20ns_h}
        }
        \subfigure[]
        {
        \includegraphics[height=20mm]{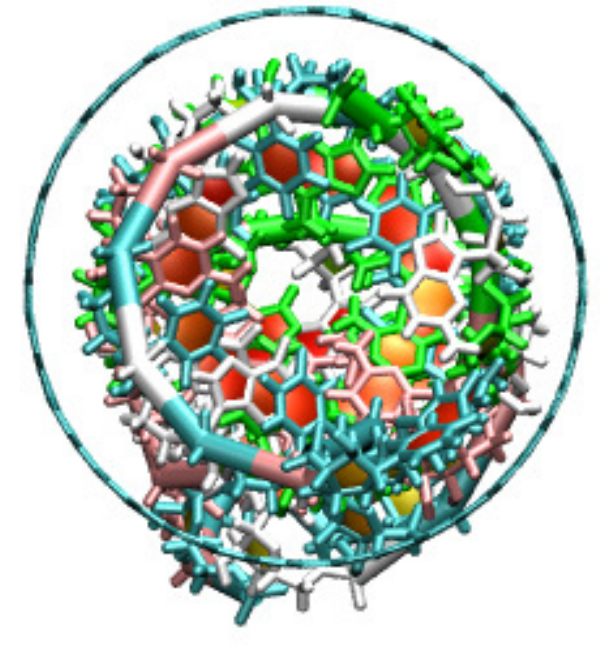}
        \label{20ns_v}
        }\\
        \subfigure[35 ns]
        {
        \includegraphics[height=20mm]{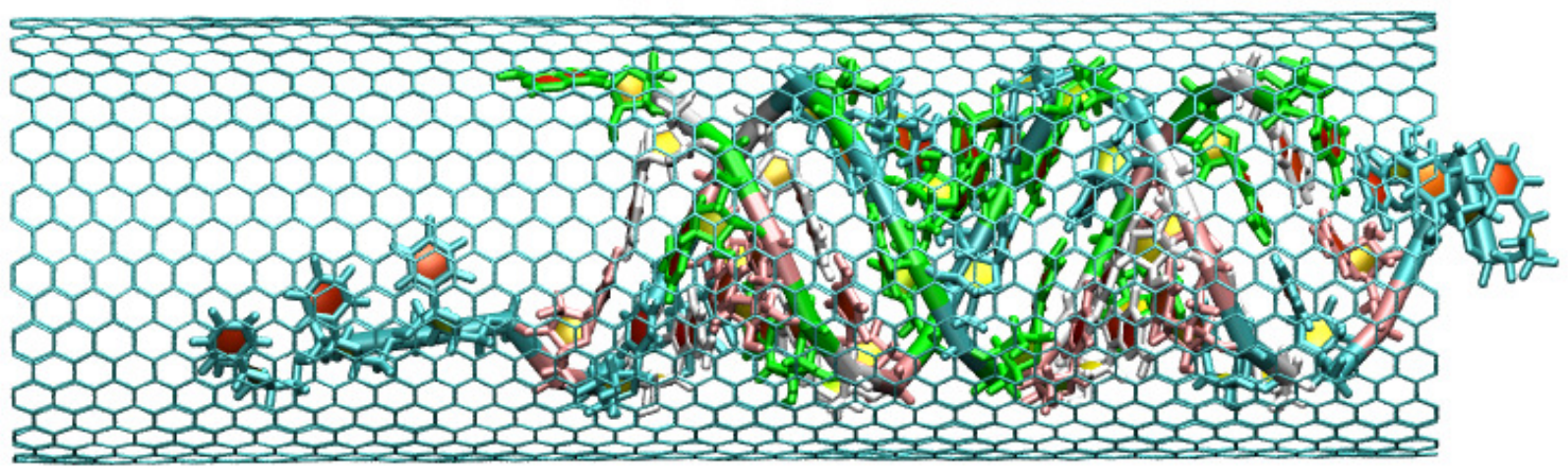}
        \label{35ns_h}
        }
        \subfigure[]
        {
        \includegraphics[height=20mm]{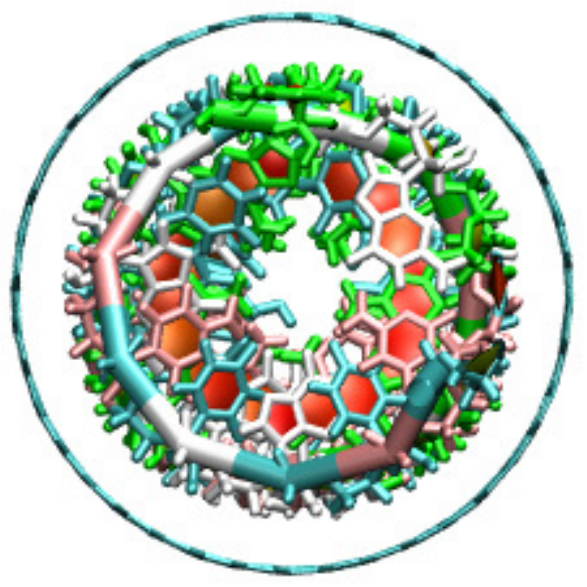}
        \label{35ns_v}
        }\\
        \caption{(a) The initial system setup 
for simulation where siRNA-CNT hybrid was 
solvated with water and neutralizing Na$^{+}$ 
counterions. (b)-(i) Snapshots of siRNA inside 
(20, 20) CNT at 3 ns, 10 ns, 20 ns and 35 ns 
during MD simulation in horizontal and vertical 
view with respect to CNT long axis $\hat{n}$. 
Water and counterions were not shown in Figures 
(b)-(i) for clarity. The snapshots in this paper 
were rendered using VMD software \cite{humphrey}.}
        \label{snapshots1}
\end{figure*}

Snapshots shown in Figure \ref{snapshots1} represent the
translocation process of siRNA inside (20, 20) CNT
at various instants of time. To present the details 
of the system studied, we show the initial 
siRNA-(20, 20) CNT system with counterions and water
in Figure \ref{sirna-ini}. In Figures \ref{3ns_h} to 
\ref{35ns_v}, we show the instantaneous snapshots 
of the system in few ns interval. For clarity, water 
and counterions are not shown.
From the snapshots and {subsequent} analysis, we observe that
the siRNA translocates into the interior of (20, 20) CNT 
from the initial configuration. The translocation 
of siRNA inside (20, 20) CNT is very fast and 
happens in 17.4 ns. After the translocation, 
siRNA stays inside CNT with least positional 
fluctuations during the remaining simulation period of 30-40 ns.
We have also observed the translocation
of siRNA at 107 mM, inside (19, 19), (18, 18) and (25, 15) CNTs.
The siRNA inside (20, 20) CNT at 0 mM has least 
structural deformations compared to all other CNTs of 
different diameter and chirality studied in this work.
Similar results are observed at 107 mM NaCl concentration also.
In all these cases, the siRNA is very 
stable inside CNT without any further movement 
after the translocation. This siRNA-CNT hybrid 
can serve as siRNA delivery vector for the 
delivery of siRNA for RNAi therapy.
For the siRNA release after delivery, functionalizing 
siRNA with nanoparticles or application of electric 
field can be used.\\

\begin{figure*}
        \centering
        {
        \begin{overpic}[height=60mm]{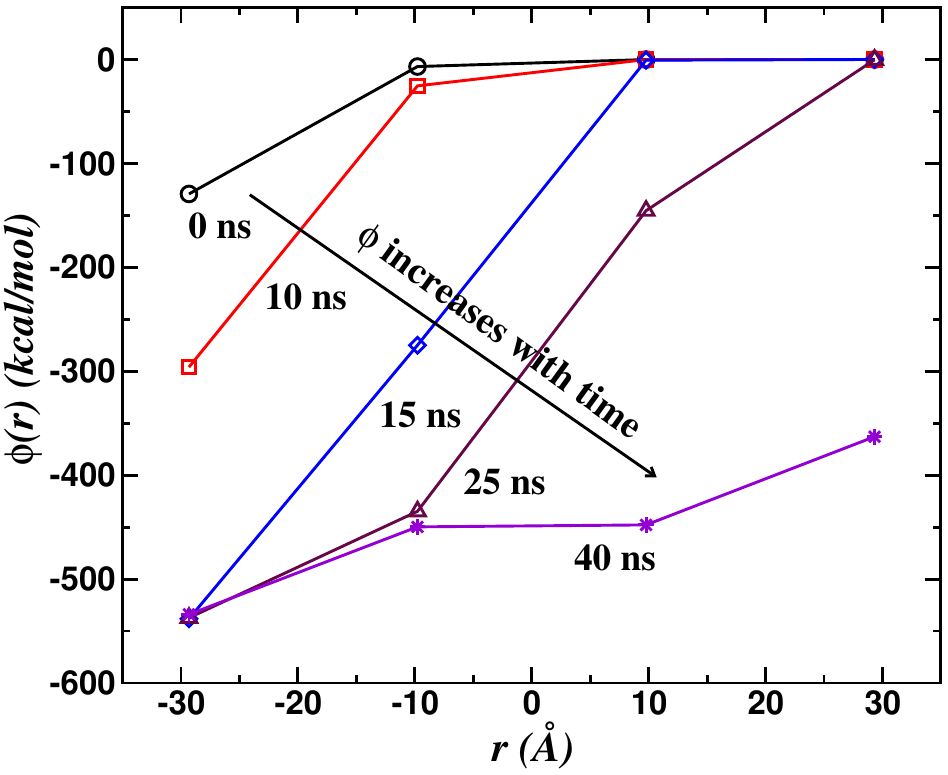}
        \end{overpic}
        }
\caption{The van der Waals interaction energy 
$\phi\left(r\right)$ of siRNA that is arising due 
to the presence of CNT in the system. At $t$ = 0 ns,
siRNA is outside CNT with two sticky-ends lying 
near to one end of the CNT. Hence at $t$ = 0 ns,
the interaction of siRNA with CNT is very less. 
As time progress these sticky-ends would interact 
with CNT strongly as shown at $t$ = 10, 15, 25 and 40 ns.}
\label{vdw}
\end{figure*}

To understand the energetics of siRNA as it translocates 
inside (20, 20) CNT, we have calculated {the} van der Waals 
(vdW) interaction energy $\phi(r)$ between siRNA and CNT at various 
instants of time. We plot $\phi(r)$ as a function of the 
siRNA spatial position $r$ at various instants of time 
$t$ in Figure \ref{vdw}. Figure \ref{vdw} shows how different 
parts of siRNA interact with CNT and also the interaction strength 
along CNT axis $\hat{n}$. In Figure \ref{vdw}, $r=0$ 
corresponds to the center of mass (COM) position of siRNA.
For the calculation of $\phi(r)$, siRNA has been divided
into 4 parts along its helix axis. The data shown in 
Figure \ref{vdw} represents how the COM of these 4 parts of 
siRNA interact with CNT as it translocates. 
It can be seen from the Figure \ref{vdw} that $\phi(r)$ has 
different shapes at $t$ = 0 ns, 10 ns, 15 ns, 25 ns 
and 40 ns. Initially siRNA is outside CNT with only two sticky-ends
{(these are un-paired nucleobases in a nucleic acid)}
and one intact Watson-Crick (WC) base-pair lying inside CNT. 
Hence the interaction of siRNA with CNT is very less which 
are far apart from each other (see $t=0$ ns curve in Figure \ref{vdw}).
As $t$ increases, these sticky-ends interact strongly with 
CNT as shown in Figure \ref{vdw} at $t$ = 10 ns, 15 ns, 25 ns 
and 40 ns. The siRNA strongly interacts with CNT after 
complete translocation for $t\ge \tau$; where $\tau$
is the translocation time. The increase of $\phi$ after
complete translocation is about 407 kcal/mol or 690.8 $k_BT$.

\subsection{Free energy landscape during translocation}
\label{freeenergyeqmd}
\subsubsection{Free energy using equilibrium simulation: MM-GBSA and 2PT method}
In general, the binding free energy for the non-covalent 
association of two molecules {A and B} may be written as
$\Delta{G}\left(A+B\rightarrow{AB}\right)=G_{AB}-G_{A}-G_{B}$.
For any species on the right hand side $G(X)=H(X)-TS(X)${, where
$H$ is the enthalpy, $T$ is the absolute temperature and $S$ is the entropy of the molecule.
Therefore} the binding energy at constant temperature can be written as
\begin{equation}
\Delta{G}=\Delta{H}-T\Delta{S}
\label{deltag}
\end{equation}
The calculation of enthalpy difference ($\Delta{H}$) 
and entropy difference ($\Delta{S}$) contributions 
to the binding free energy in Eqn. \ref{deltag} are done 
using molecular mechanics - generalized Born surface 
area (MM-GBSA) method and 2 phase thermodynamic (2PT) 
method \cite{lin2003,lin2010,pascal2011}, respectively.
{The change in enthalpy $\Delta{H}$
can be decomposed into the gas-phase energy 
$\left(\Delta{E}_{\text{gas}}\right)$ and solvation 
free energy $\left(\Delta{G}_{\text{sol}}\right)$; 
{\it i.e.,} $\Delta{H}=\Delta{E}_{\text{gas}}+\Delta{G}_{\text{sol}}$.
For any species $A, B$ or complex $AB$, the gas-phase energy, $E_{\text{gas}}$ is calculated from
molecular mechanics using $E_{\text{gas}}=E_{\text{ele}}+E_{\text{vdw}}+E_{\text{int}}$,
where, $E_{\text{ele}}$ is the electrostatic energy calculated from the Coulomb 
potential, $E_{\text{vdw}}$ is the non-bonded van der Waals energy and
$E_{\text{int}}$ is the internal energy contribution arising from bond 
stretching, angle bending and dihedrals.
On the other hand, the solvation free energy, $G_{\text{sol}}$
can be further decomposed as electrostatic $G_{\text{el}}$ and 
non-electrostatic $G_{\text{non-el}}$ contributions; {\it i.e.,} 
$G_{\text{sol}}=G_{\text{el}}+G_{\text{non-el}}$. The electrostatic 
energy, $G_{\text{el}}$ is calculated from Generalized Born (GB) 
method which assumes that the atoms in a molecule are spheres 
of radius $R_i$ (called Born radius) and have an effective charge $q_i$
(called Born charge). The molecule is assumed to be surrounded
by a solvent of dielectric constant $\epsilon$ (80 for water at 300 K)
and the solute atoms have a dielectric constant of 1.
The analytic expression for the $G_{\text{el}}$ in the 
GB model \cite{still1990,srinivasan1999} is given by
\begin{align}
G_{\text{el}} = -\frac{1}{2}\sum_{i,j}\frac{q_iq_j}{f_{GB}(r_{ij}, R_i, R_j)} \left(1-\frac{\exp{\left(-\kappa f_{GB}\right)}}{\epsilon}\right)
\end{align}
where 
\begin{align}
f_{GB}(r_{ij}, R_i, R_j) = \left[r_{ij}^2 + R_i R_j \exp{\left(-r_{ij}^2/4R_i R_j\right)}\right]^{\frac{1}{2}}
\end{align}
where $r_{ij}$ is the distance between atoms $i$, $j$ and
$R_i$ are effective Born radii and $\kappa$ is the 
Debye-H\"{u}ckel screening parameter.
The non-electrostatic energy, $G_{\text{non-el}}$ is calculated as
$\gamma\times{SASA}+\beta$; where $\gamma$ is the
surface tension parameter ($\gamma$ = 0.0072
kcal/mol-\AA$^2$; $\beta$ = 0 kcal/mol) and $SASA$ is the solvent-accessible
surface area of the molecule.
The entropy contribution appearing in Eqn. \ref{deltag}
is carried out using 2 phase thermodynamic approach proposed
by Lin {\it et. al.} \cite{lin2003,lin2010,pascal2011}
which is motivated by the observation that the density of states ($S(\nu)$)
of a liquid {can be decomposed into a gas 
component and a solid component} \cite{lin2003,lin2010,pascal2011}.
The density of states function can be calculated from the 
Fourier transform of velocity auto-correlation function; {\it i.e.,}
\begin{align}
S(\nu) = \frac{2}{k_BT}\lim_{\tau\to\infty} \int_{-\tau}^{\tau} \sum_{j=1}^{N} \sum_{k=1}^{3} m_j {c_j}^k(t) e^{-i2\pi v t} dt
\end{align}
where ${c_j}^k(t)$ is the velocity auto-correlation function
which is given by
\begin{align}
{c_j}^k(t) = \lim_{\tau\to\infty} \frac{1}{2\tau} \int_{-\tau}^{\tau} {v_j}^k(t) {v_j}^k(t+t^{\prime}) dt^{\prime}.
\end{align}
The entropy $S$ is then calculated from the knowledge of $S(\nu)$
using
\begin{align}
S = k_B {\text{\normalsize ln}}Q + \beta^{-1} {\frac{\partial {\text{\normalsize ln}}Q}{\partial T}}_{N,V}
\end{align}
where the partition function $Q$ is given by
\begin{align}
{\text{\normalsize ln}} Q = \int_{0}^{\infty} d\nu S(\nu) {\text{\normalsize ln}q_{HO}(\nu)}.
\end{align}
The decomposition of $S(\nu)$ into
gas-phase and solid-phase is performed to get the entropy of
molecular species $A$, $B$ or complex $AB$ to get finally
the change in free energy $T \Delta{S}$. 
The 2PT method }has found successful application in several related 
problems \cite{lin2003,maitinl,lin2010,pascal2011,hemant,nandy2011}.\\

\begin{figure*}
        \centering
        \subfigure[]
        {
        \includegraphics[height=60mm]{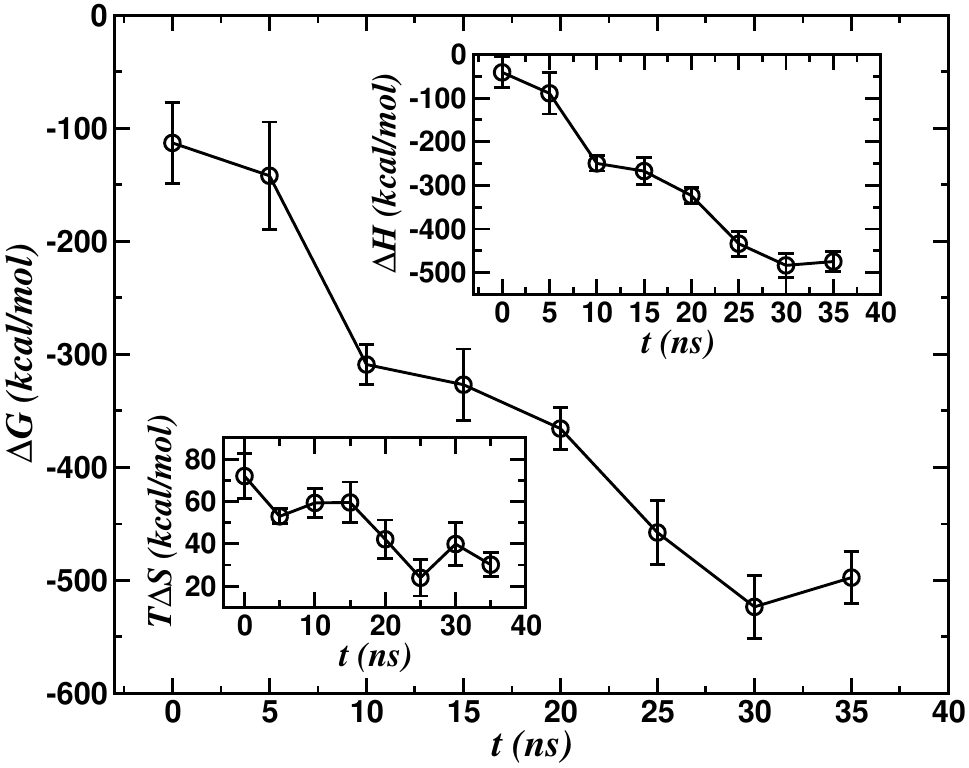}
        \label{free_energy_time}
        }
        \subfigure[]
        {
        \begin{overpic}[height=6cm]{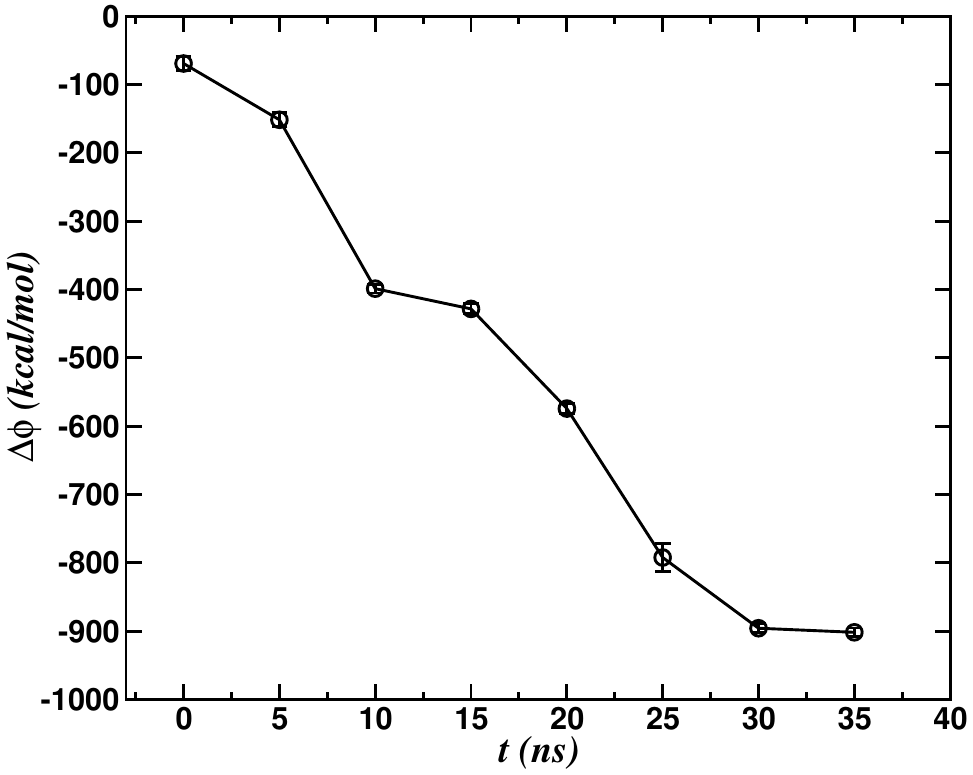}
        \put(30,68){\includegraphics[height=0.55cm]{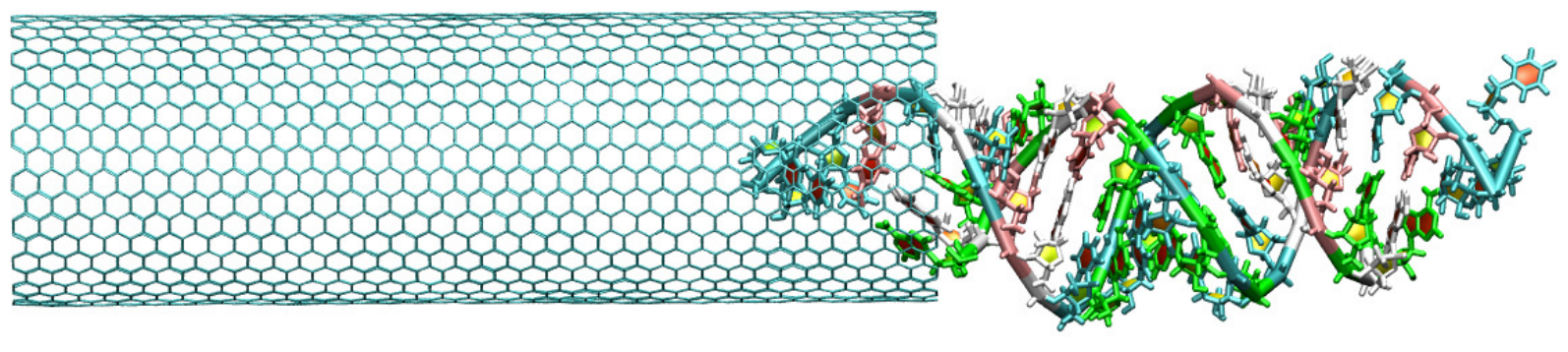}}
        \put(52,45){\includegraphics[height=0.55cm]{20ns_horizontal}}
        \put(72,20){\includegraphics[height=0.55cm]{35ns_horizontal}}
        \end{overpic}
        \label{vdw-time}
        }
\caption{(a) Binding free energy as a function of time with 
entropy and enthalpy shown in insets that are calculated
using 2PT and MM-GBSA method, respectively. (b) vdW
contribution to $\Delta{G}$ as a function of time.
$\Delta{G}$ and $\phi$ are increasing with time as
siRNA translocates inside (20, 20) CNT and reaches
constant after complete translocation.}
\label{2pt-mmpbsa}
\end{figure*}

The binding free energy ($\Delta{G}$) as function of 
time is shown in Figure \ref{free_energy_time}. 
Entropy ($T\Delta{S}$) and enthalpy ($\Delta{H}$) 
contributions to $\Delta{G}$ are shown in the inset. 
$\Delta{S}$ decreases with time 
since the fluctuations in siRNA nucleobases are 
suppressed by CNT making less microstates available 
for siRNA as translocation progresses. However,
$\Delta{H}$ is increasing with time and attains constant value
after complete translocation. $\Delta{G}$
also follows similar trend since the entropy contribution
is very small compared to enthalpy contribution.
We plot the vdW contribution $\Delta{\phi}$
to the total binding energy in Figure \ref{vdw-time}. When 
siRNA and CNT are far apart in the initial stage, $\Delta{\phi}$
is very less but increases as siRNA translocates inside CNT.
We also show instantaneous snapshots of the siRNA-CNT 
hybrid at various instants of time $t$ in Figure \ref{vdw-time}.
 $\Delta{\phi}$ is -900 kcal/mol after complete translocation of siRNA 
compared to -85 kcal/mol at the initial stage. 
The interaction between siRNA and CNT is driven 
mainly by vdW interaction (Figure \ref{vdw-time}).
This is similar to our earlier results that the unzipping and adsorption
of siRNA on CNT/graphene is also driven by the
vdW interaction \cite{santoshjcp,santoshgraphene}.
{The conversion of time into distance 
between CNT and siRNA while plotting $\Delta G(r)$ has 
been discussed in the supplementary materials \cite{supplementary}}.

\subsubsection{Free energy using Umbrella sampling method}
{In the un-biased MD simulations discussed so far, siRNA has 
translational motion in 1-D along nanotube axis $\hat{n}$
in order to translocate into the interior of CNT.
The free energy landscape of the siRNA 
translocation process along such suitable 
reaction coordinate will help in understanding
the stable and unstable states for siRNA 
with respect to the CNT. In the previous 
section we have described the free energy calculation 
from the equilibrium un-biased MD simulation
using a combination of MM-GBSA and 2PT methods. 
Such methods in equilibrium MD cannot properly 
account for the sampling of high energy 
states. For this reason, we use Umbrella sampling 
method \cite{torrie} to sample the entire phase 
space along chosen reaction coordinate and calculate the free 
energy profile for siRNA translocation inside CNT.
The basic idea implemented in constructing the 
free energy is to add a biasing potential
to obtain sampling of less probable states
for siRNA translocation process and record
the biased histograms. This is done for a
series of biasing potentials at various values of chosen
reaction coordinate that span the entire
translocation path of interest. We make
sure that the successive histograms
have enough overlap in order to reconstruct
the un-biased free energy from biased histograms.
Below we give a brief overview of the Umbrella 
sampling method. For further details, readers 
are refereed to the excellent text book by 
Frenkel and Smit \cite{ums}.}\\

{With the unperturbed potential $V(\mathbf{r}^{N})$,
we have added a biasing harmonic potential
$U_i(\xi) = \frac{1}{2}k\left(\xi-\xi_i\right)^2$,
resulting in the perturbed potential
$V^{\prime}(\mathbf{r}^{N}) = V(\mathbf{r}^{N}) + \sum_{i=1}^{N_w}{U_i(\xi)}$;
where $\xi_i$ is the restrained distance between
the center of masses of `far end' of CNT and first two Watson-Crick
hydrogen bonded base-pairs of siRNA close to the CNT for
$i^{\text{th}}$ window. The values of $k$ and $\Delta\xi$
should be chosen optimally such that the phase space
is properly sampled within reasonable time scales.
Therefore, we have optimized the force constant $k$
of $U_i(\xi)$, $\Delta\xi$ and equilibration
time before performing the Umbrella
sampling. The optimized values are $k$ = 4 kcal/mol-\AA$^2$,
$\Delta\xi_i$ = 1 \AA~ and equilibration time of 1 ns to ensure
proper sampling and overlapping of successive histograms,
 ${P_i}^{\prime}(\xi)$. The values of $\xi_i$ is changed
from 130 \AA~ to 10 \AA~ until the complete translocation
happens in steps of 1 \AA~ totaling $N_w$ = 120 Umbrella
simulations  (windows) of each 1 nano second duration.
The probability distribution of the reaction coordinate
$\xi$ separating CNT and siRNA is,
\begin{eqnarray*}
P_i(\xi) &=& \langle\delta\left(\xi-\xi_i\right)\rangle \\
     &=& \frac{1}{Z}\int{d\mathbf{r}^{N} \delta\left(\xi-\xi_i\right) \text{exp} \left[-\beta{V}(\mathbf{r}^{N})\right]}
\end{eqnarray*}
Similarly the probability distribution in the presence of $U_i(\xi)$ is
\begin{eqnarray*}
{P_i}^{\prime}(\xi) &=& {\frac{1}{Z^{\prime}}\int{d\mathbf{r}^{N}
 \delta\left(\xi-\xi_i\right)}}{e^{-\beta \left[V(\mathbf{r}^{N})+{U_i}(\xi)\right]}}\\
        &=& {\frac{Z}{Z^{\prime}}} {e^{-\beta {U_i}(\xi)}} P_i(\xi).
\end{eqnarray*}
Re-arranging this gives,
\begin{equation}
\label{pofi}
{P_i}(\xi) = {\frac{Z^{\prime}}{Z}} {e^{\beta {U_i}(\xi)}} {P_i}^{\prime}(\xi)
\end{equation}
here, $\delta$ is a Dirac delta function, $N$ is the number of atoms,
 $\mathbf{r}^{N}$ denotes the set of atom coordinates, $\beta = 1/k_BT$, where
$k_B$ is the Boltzmann constant, $T$ is the absolute temperature,
$V(\mathbf{r}^{N})$ is the potential energy and $Z$, $Z^{\prime}$ are
the partition functions of un-biased and biased systems, respectively.
Here problem arises due to the fact that
$P(\xi)$ becomes exceedingly small for values of $\xi$ which give
significant contribution to the free energy. Umbrella sampling
makes use of a biasing potential to sample the region of phase space
for which $P(\xi)$ is exceedingly small.
Thus, the distribution function $P(\xi)$ can be obtained using Eqn. \ref{pofi}
(to within a multiplicative constant) from the measurement
of the biased distribution $P^{\prime}(\xi)$.
The un-biased free energy ($F$) was constructed self-consistently
using the weighted histogram analysis method (WHAM) \cite{torrie,roux,ums}
with the following Eqns. \ref{wham1} and \ref{wham2}.
\begin{equation}
\label{wham1}
P(\xi) = \frac{\sum_{i=1}^{N_w}{n_i{P_i}(\xi)}}{\sum_{j=1}^{N_w}{n_je^{-\beta\left(U_j(\xi)-F_j\right)}}}
\end{equation} 
where $F_i$ is given by 
\begin{equation}
\label{wham2}
e^{-\beta F_i} = \int d\xi e^{-\beta U_i(\xi)} P(\xi)
\end{equation}
where $n_i$ is the number of data points in $i^{\text{th}}$
window. By piecing together the relative free energies measured using a
number of biasing potentials, we construct $F(\xi)$
over the chosen range of $\xi$ = 10 \AA~ to 130 \AA.}\\

\begin{figure*}
        \centering
	\subfigure[]
	{
        \begin{overpic}[height=6.0cm]{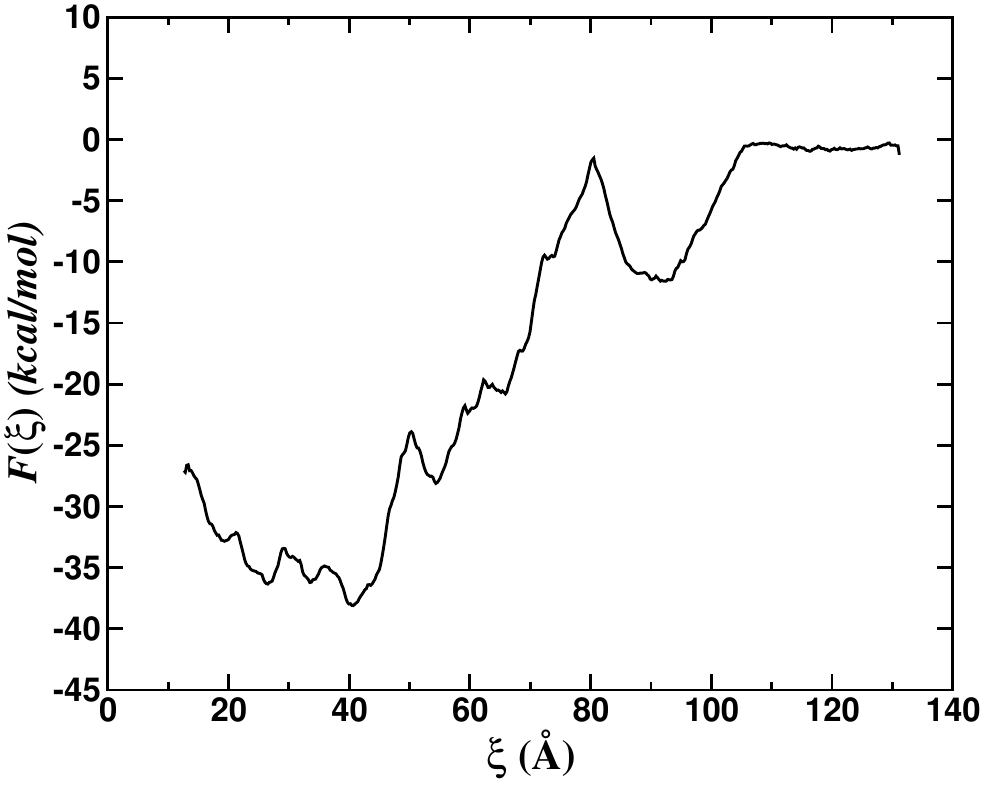}
        \put(74,55){\includegraphics[height=0.50cm]{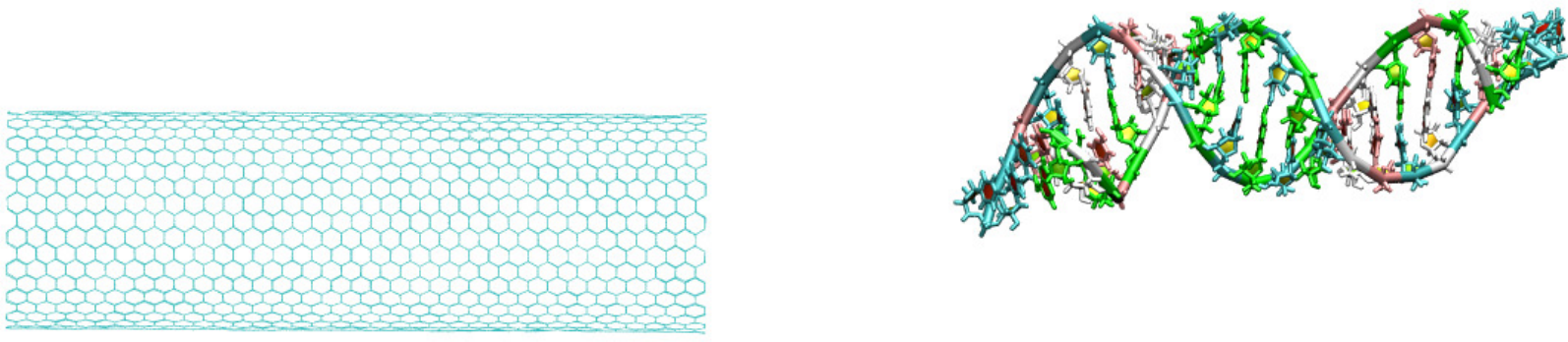}}
        \put(80,62){$\downarrow$}
        \put(47,69){\includegraphics[height=0.5cm]{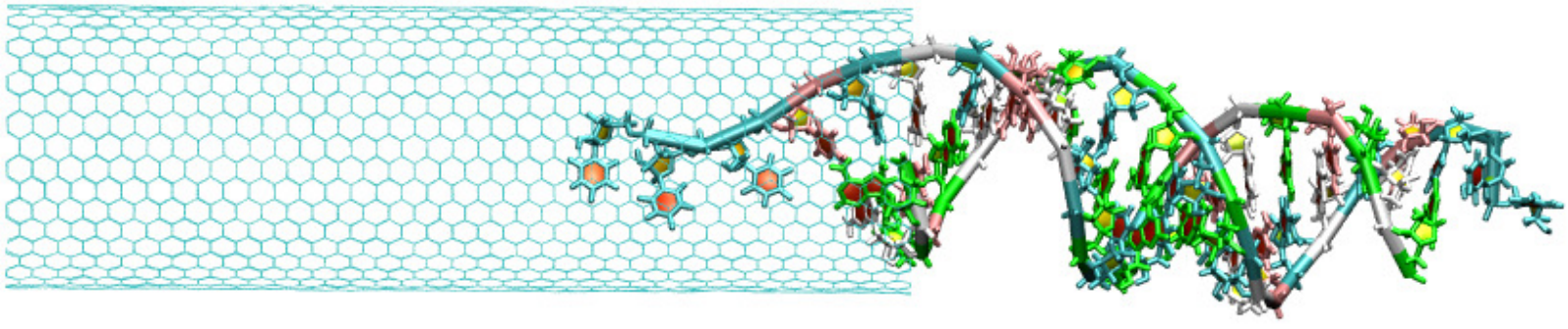}}
        \put(58.8,65){$\uparrow$}
        \put(60,41){\includegraphics[height=0.5cm]{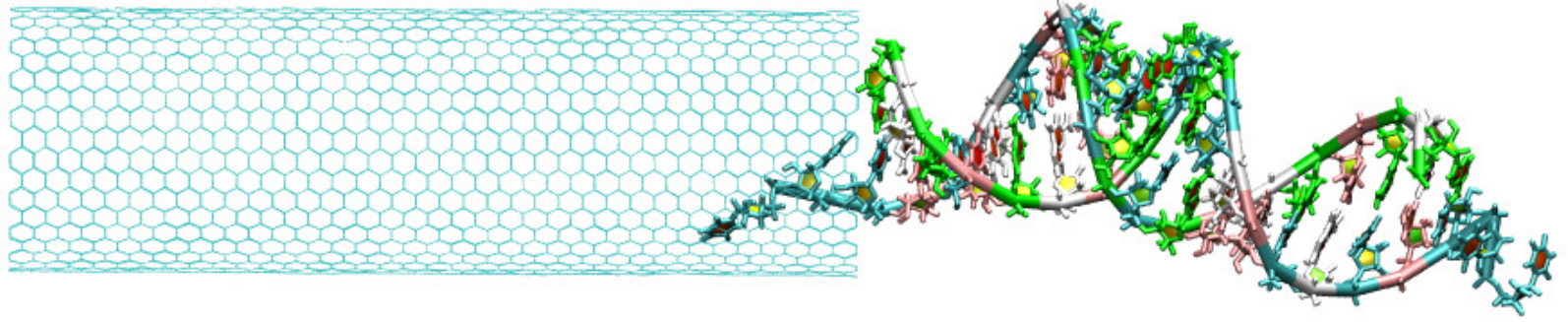}}
        \put(66,50){$\downarrow$}
        \put(53,29){\includegraphics[height=0.5cm]{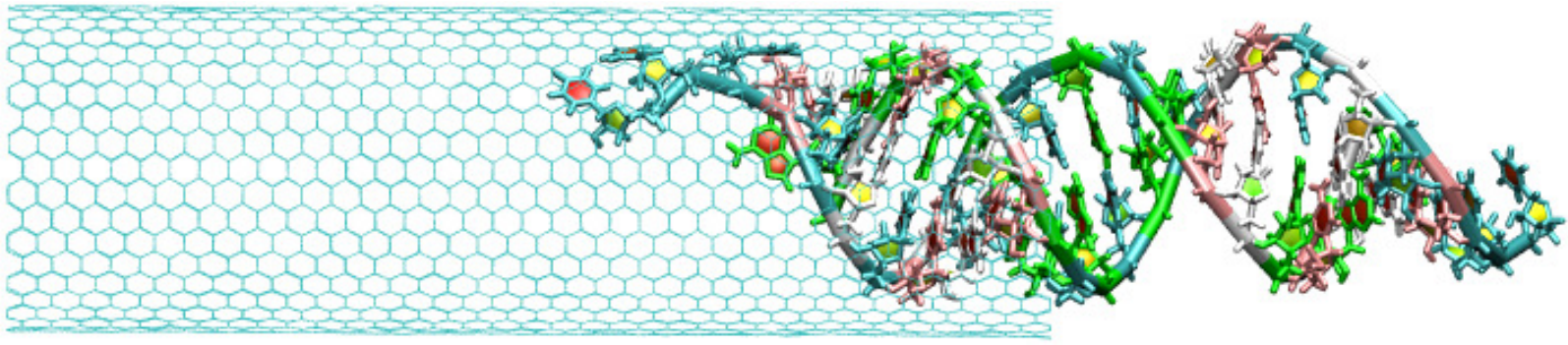}}
        \put(49.5,38){$\searrow$}
        \put(42,13){\includegraphics[height=0.5cm]{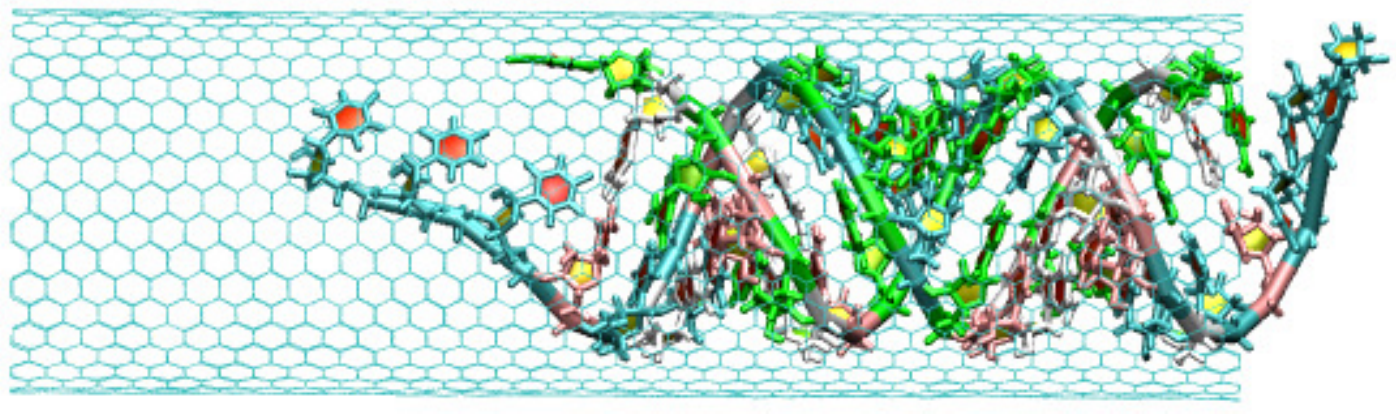}}
        \put(35,16.5){$\rightarrow$}
        \end{overpic}
	\label{pmf_sirna_ff10}
	}
	\subfigure[]
	{
	\includegraphics[height=6.0cm]{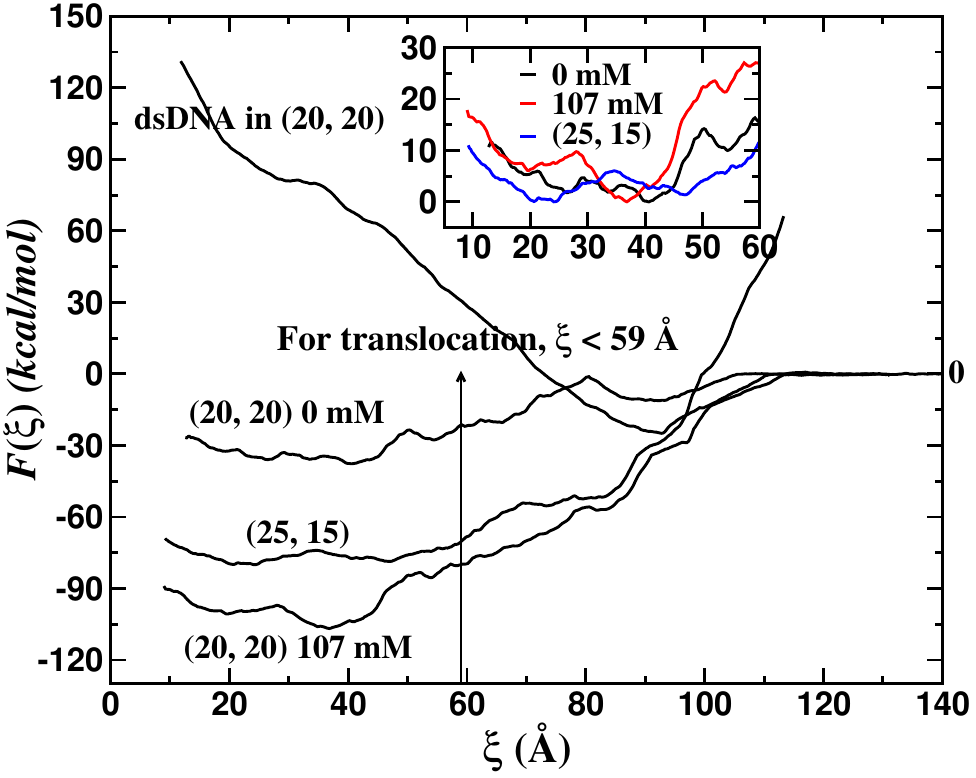}
	\label{pmf_all}
	}
\caption{(a) Potential of the mean force $F(\xi)$ of 
siRNA translocation inside (20, 20) CNT along
the reaction coordinate $\xi$. (b) $F(\xi)$ for siRNA
translocation inside (20, 20) CNT at 0 and 107 mM NaCl,
(25, 15) CNT and dsDNA translocation inside (20, 20) 
CNT. We define that the siRNA/dsDNA is translocated 
inside CNT when at least half of the base-pairs of
siRNA/dsDNA are inside CNT from which it follows 
that $\xi \le$ 59 \AA. The inset shows the minima
of shifted $F(\xi)$ which is inside CNT for siRNA and
outside for dsDNA since it does not translocate.}
\label{pmf}
\end{figure*}

Figure \ref{pmf_sirna_ff10} shows the free energy $F(\xi)$ for siRNA 
in (20, 20) CNT at 0 mM NaCl concentration.
We find that when siRNA is outside (20, 20)
CNT, the free energy is zero since there is 
no interaction between siRNA and (20, 20) CNT.
The snapshots at various time instants show the translocation
process as $\xi$ is decreasing from 130 \AA~ to 10 \AA.
For $\xi \ge$ 105 \AA, siRNA is outside (20, 20) CNT and 
does not interact with it. As the value of $\xi$ is
decreased from 105 \AA, $F(\xi)$ decreases to
minima at 92 \AA~ at which siRNA is in a local
stable state. At this stage siRNA rotates inside 
CNT and two unpaired bases of siRNA and one WC 
base-pair unzips to interact with (20, 20) CNT 
as can be seen in Figure \ref{pmf_sirna_ff10}.
Further decrease in $\xi$ results in increasing 
$F(\xi)$ having an energy barrier for siRNA to 
translocate inside CNT. $F(\xi)$ reaches maximum 
at $\xi$ = 80 \AA~ with an energy barrier 
of $\Delta{G}$ = 9 kcal/mol.
The energy barrier for further translocation 
arises due to strong vdW interaction between 
unpaired nucleobases of siRNA and CNT. 
{Thermal fluctuations help siRNA 
to overcome this barrier}. Note that, once it overcomes 
this barrier, favorable vdW interaction between 
nucleobases and CNT helps siRNA to translocate inside. 
From $\xi$ = 80 \AA, siRNA goes inside (20, 20) CNT
easily and $F(\xi)$ decreases until complete siRNA 
translocation at $\xi_{\text{min}}$ = 42 \AA. 
This is the most stable position for siRNA inside (20, 20) CNT 
where siRNA stays for the rest of the simulation time.\\

Figure \ref{pmf_all} presents free energy 
for siRNA in CNTs for (20, 20), (25, 15) chiralities
at 0 mM NaCl concentration, for (20, 20) CNT
at 107 mM NaCl concentration and for dsDNA in 
(20, 20) CNT at 0 mM NaCl concentration.
In the un-biased simulation, we have observed 
the translocation of siRNA in CNTs but no translocation 
of dsDNA is observed. The dsDNA has favorable 
state only outside the (20, 20) CNT as can be 
seen in Figure \ref{pmf_all}. This might be due 
to the less favorable interaction of thymidine
nucleobase with nanotube compared to the uridine
nucleobase interaction with nanotube \cite{santoshjcp,santoshgraphene}.
For $\xi \le {L}/{2}$, where $L$ is the length
of the CNT, $F(\xi)$ is constantly increasing
which means dsDNA encounters unfavorable free 
energy landscape when translocated inside 
(20, 20) CNT. This dramatic difference of 
interaction between CNT and siRNA and CNT 
and dsDNA may be due to the relatively 
weaker interaction strength of thymidine
with CNT than that of uridine with CNT \cite{santoshjcp,santoshgraphene}.
Moreover, the stronger WC base-pairing interaction 
energy of A-T compared to A-U \cite{sponer2004,
santoshjcp,santoshgraphene,huang2011} makes
dsDNA difficult to get unzipped. Earlier we 
have shown that unzipped base-pairs facilitate 
binding with CNT/graphene \cite{santoshjcp,santoshgraphene}. dsDNA 
requires more than 100 kcal/mol energy to 
overcome a free energy barrier to translocate 
inside (20, 20) CNT which is not possible 
without any external force. However, possibility 
of the translocation of dsDNA in CNT of large 
diameter has not been studied in our simulation. 
In experiments of dsDNA translocation inside 
CNT, the diameter of CNT is 50-100 nm 
\cite{ito2003,fan2005nl}.
In the experiments of ssDNA translocation 
through $\alpha$-hemolysin ion 
channel \cite{kasianowicz1996}, the diameter of 
$\alpha$-hemolysin is 26 \AA~ and the polymer is 
single stranded. Diameter of (20, 20) CNT is 26.74 
\AA~ which is close to the diameter of $\alpha$-hemolysin 
ion channel and so the channel diameter is not sufficient 
for dsDNA translocation. We have observed the 
translocation of ssDNA through CNT of diameter 26.74 \AA~ 
\cite{santoshjbs}. Our results on ssDNA and dsDNA 
translocation are consistent with experiments \cite{kasianowicz1996}.
$F(\xi)$ for dsDNA has global minimum only outside 
the (20, 20) CNT. When dsDNA is forced to translocate 
inside (20, 20) CNT with an external Umbrella
potential, dsDNA deforms largely by breaking 
most of the WC base-pairs due to the large external 
force by Umbrella potential. To have further 
confirmation that siRNA translocate inside CNT 
and dsDNA does not, we have performed two 
separate simulations of siRNA and dsDNA by 
keeping them initially inside the (20, 20) CNT.
As expected, siRNA stays inside the CNT for 
long time where as dsDNA comes out of the CNT 
within 500 ps. These results confirm that 
siRNA spontaneously translocates inside a 
(20, 20) CNT without any external force 
where as dsDNA requires an external force 
in order to be translocated inside (20, 20) CNT.\\ 

To test the effect of chirality as well 
as salt concentration on the translocation 
event, we have calculated the free energy
profile of siRNA translocation inside (25, 15)
CNT at 0 mM NaCl concentration and for (20, 20)
CNT at 107 mM NaCl concentration using Umbrella 
sampling method. The free energy profile 
for these cases are shown in Figure \ref{pmf_all}.
In both these cases, the free energy minima is lower 
compared to the case of siRNA in (20, 20) CNT at 
0 mM concentration of NaCl. So for these
cases we expect translocation to happen faster 
compared to the case for siRNA in (20, 20) CNT at 0
mM NaCl concentration. Translocation time for various 
cases has been discussed in section \ref{section_transtime}. The inset of
Figure \ref{pmf_all} shows the minima of $F(\xi)$. 
We observe that $\xi_{\text{min}}$ = 37 \AA~ at 107 
mM as compared to $\xi_{\text{min}}$ = 42 \AA~ at 0 mM.
However, $\xi_{\text{min}}$ = 21 \AA~ for siRNA in
(25, 15) CNT. The origin for this observation is 
discussed in the next section by analyzing 
un-biased simulations.

\subsection{Translocation time ($\tau$)}
\label{section_transtime}
\begin{figure*}
        \centering
        \subfigure[]
        {
        \includegraphics[height=60mm]{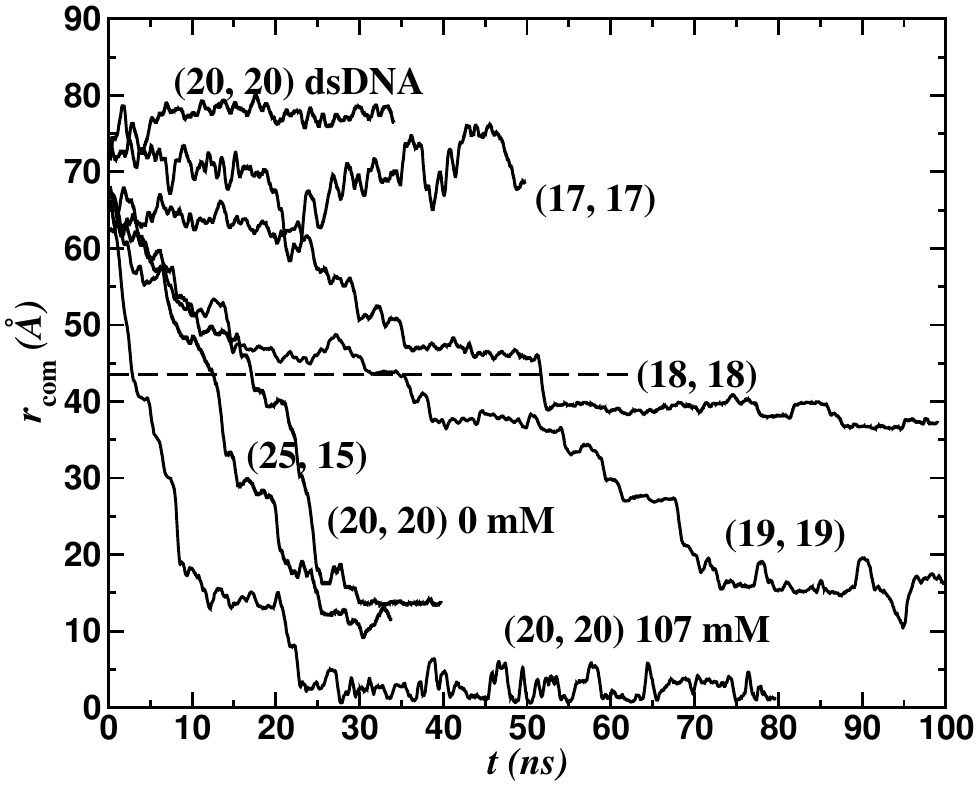}
        \label{time_com}
        }
        \subfigure[]
        {
        \includegraphics[height=60mm]{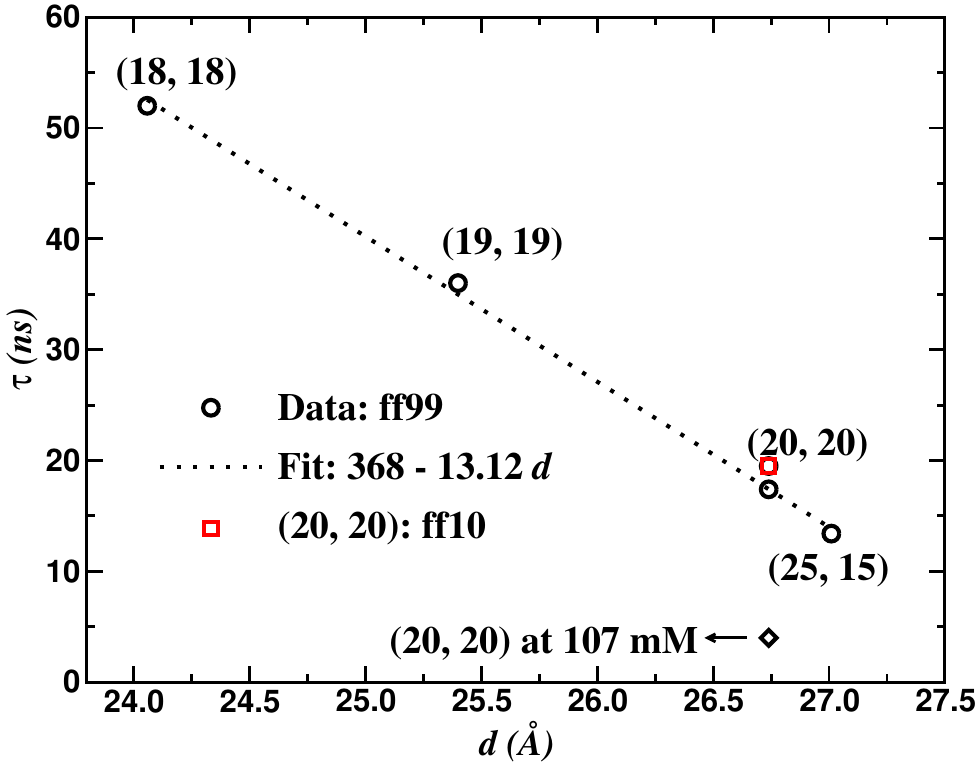}
        \label{tau}
        }
        \subfigure[]
        {
        \includegraphics[height=60mm]{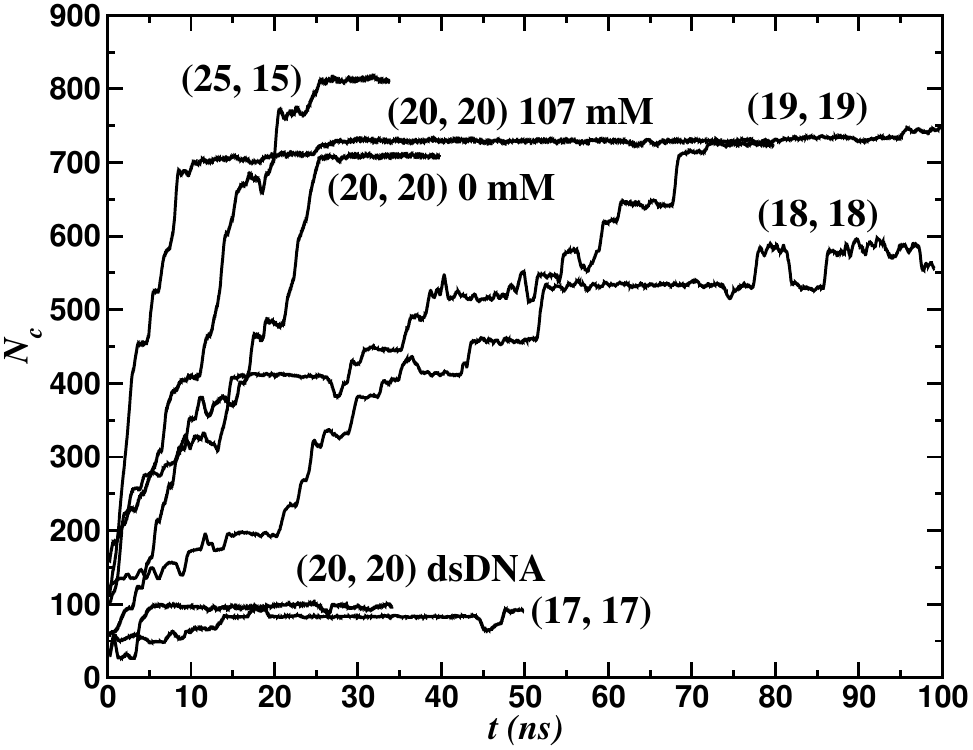}
        \label{cc}
        }
\caption{(a) The COM distance between siRNA and CNT
($r_{\text{com}}$) as shown in inset of supplementary
Figure 1(a) for siRNA inside CNT of various diameter and chirality. 
For the translocation, $r_{\text{com}} \le L/2$ implies 
$r_{\text{com}} \le 43.5$ \AA~which is marked as horizontal
dashed line. (b) Translocation time ($\tau$) of siRNA as a function 
of the diameter $d$ of CNT using ff99 force field. $\tau$~decreases 
linearly with increasing $d$. In the Figure, we have also shown
$\tau$ for siRNA in (20, 20) CNT using ff10 force field as red 
square mark and siRNA in (20, 20) CNT at 107 mM as diamond mark.
(c) Number of close contacts ($N_c$)}
\label{com-tau}
\end{figure*}

The translocation of siRNA strongly depends on the
diameter $d$ of CNT since the vdW interaction between 
siRNA and CNT increases as $d$ is decreased (nearly 
comparable to siRNA diameter). The translocation time,
$\tau$ is defined
as the time required for at least half of the siRNA 
base-pairs to enter inside the CNT; {\it{i.e.,}} $r_{\text{com}}
\le L/2$ which corresponds to $r_{\text{com}} \le 43.5$ \AA~ 
or $\xi \le 59$ \AA, where $L$ = 87 \AA~ is the length of the CNT. 
We plot $r_{\text{com}}$ and $\tau$
in Figures \ref{time_com} and \ref{tau}, 
respectively for various diameters
of CNT. In Figure \ref{time_com}, we see 
that $r_{\text{com}}$ is decreasing with time as 
siRNA moves towards the interior of the CNT. 
However, the sticky-ends and unzipped 
nucleobases of siRNA interact with CNT 
in the course of translocation, and 
these give rise to very rugged free 
energy profile of siRNA during the 
translocation. We {have} calculated $\tau$ 
by tracking $r_{\text{com}}$ as a function 
of time as shown in Figure \ref{time_com}.
The dashed horizontal curve drawn at $L/2$
serves as the reference for translocation criteria.
We find that $\tau$ is linearly decreasing 
with CNT diameter $d$. 
Similar dependence was observed for DNA transport 
through graphene nanopore \cite{garaj} where 
the ionic conductance {was found to be} proportional to the 
pore diameter. We note that our results on $\tau$
are very sensitive to the initial model building.
{To see the effect of initial relative positioning 
of siRNA with respect to CNT, we simulated the translocation                   
process for three different initial conditions for (20, 20) CNT. 
We find $\tau$ to be 18 ns, 14 ns and 17.4 ns and see
a strong dependence of $\tau$ on the initial conditions.}
Hence our results on $\tau$ can only serve as
qualitative understanding of the
systems under investigation in this study.
To understand the origin of this diameter 
dependence, we have calculated the number 
of contacts $N_c$ between the siRNA and CNTs 
and have shown them in Figure \ref{cc}.
$N_c$ of siRNA is calculated within 5 \AA 
from the inner surface of CNT that mostly 
represents the effective vdW interaction range. 
As shown in Figure \ref{cc}, $N_c$ of siRNA 
are 110 at $t$ = 0 ns and rapidly increases 
to a constant value after complete translocation. 
The maximum value of $N_c$ increases with $d$ 
ranging from 520 inside the (18, 18) CNT to 815 
inside the (25, 15) CNT. siRNA translocation happens
above a critical CNT diameter of 24.0 \AA~ which 
corresponds to (18, 18) CNT, below which no 
translocation is observed.
For siRNA in (17, 17) CNT and dsDNA in (20, 20) 
CNT, the value of $N_c$ is constant and very 
less indicating no translocation in the 
simulated time scale. Since we have not 
observed the translocation of siRNA in 
(17, 17) CNT and dsDNA in (20, 20) CNT, $\tau$ is
assumed to be very large. For the translocation of siRNA
inside CNT, the minimum diameter of CNT, $d_{\text{min}}$
should be greater than or equal to the diameter of siRNA 
+ effective vdW radius of CNT and siRNA;
{\it{i.e.,}} $d_{\text{min}} \ge d_{\text{siRNA}} + d_{\text{vdW}}$.
The diameter of (17, 17) CNT is 22.73 \AA~ which is 
very close to siRNA diameter and hence cannot 
accommodate siRNA unless siRNA is severely 
stretched. As expected, with decreasing CNT
diameter, the deformation in siRNA is more 
due to strong vdW interaction between siRNA 
and CNT. The structural aspects of siRNA during
translocation are discussed in section \ref{section_cc}. 
Interestingly, when the salt concentration
is increased to 107 mM, $\tau$ decreases drastically
to 4 ns compared to 17.4 ns at 0 mM salt 
concentration. High salt provides better counterion
condensation around phosphate atoms in siRNA backbone
compared to 0 mM case. The electrostatic screening 
increases as salt is increased from 0 to 107 mM which 
reduces phosphate-phosphate electrostatic repulsion in the
backbone of siRNA. Therefore, the stability to siRNA 
increases with increasing salt concentration which 
reduces the propensity of siRNA unzipping \cite{santoshjcp}.
As a result, at 107 mM NaCl concentration, 
there is less binding of the siRNA with the CNT. 
So the translocation is faster at 107 mM 
salt concentration and hence $\tau$ is small.
At 107 mM of NaCl concentration, the maximum value of 
$N_c$ is 730 which is 20 more in number than           
that of at 0 mM.

\subsection{Structural aspects of siRNA during translocation inside CNT}
\label{section_cc}
\begin{figure*}[t]
        \centering
        \subfigure[]
        {
        \includegraphics[height=60mm]{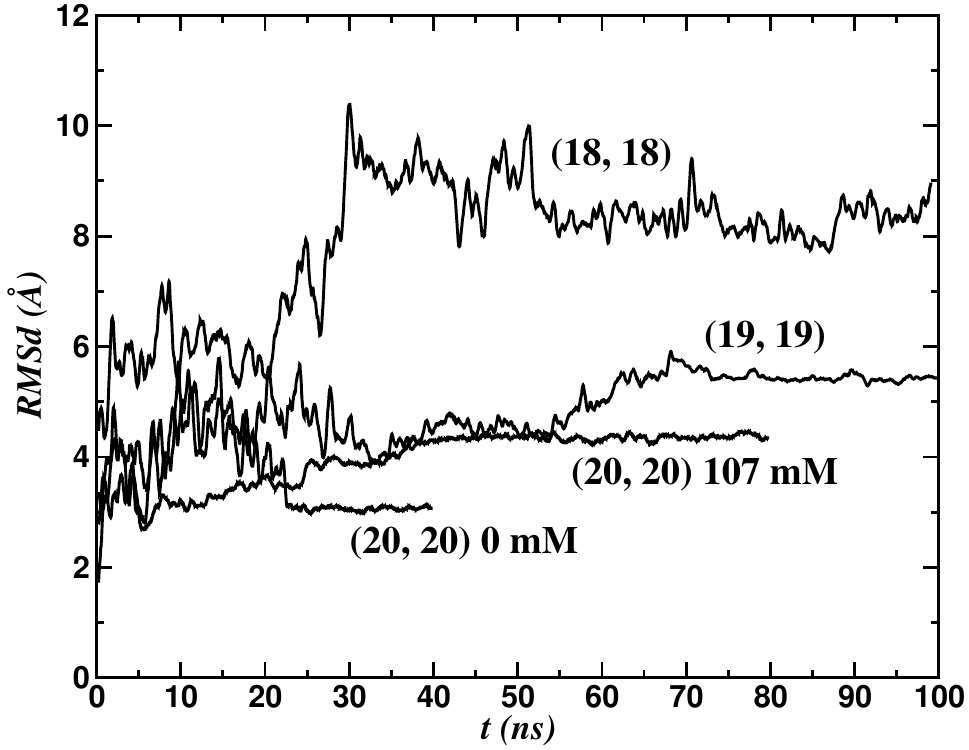}
        \label{rmsd}
        }
        \subfigure[]
        {
        \includegraphics[height=60mm]{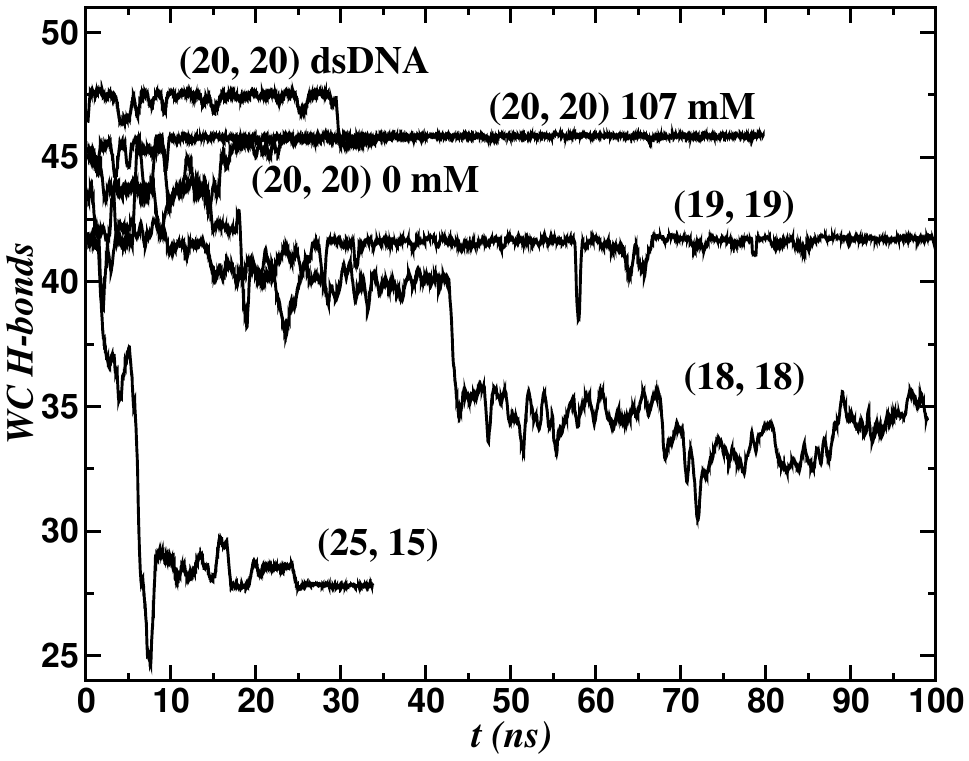}
        \label{wc}
        }
        \subfigure[]
        {
        \includegraphics[height=60mm]{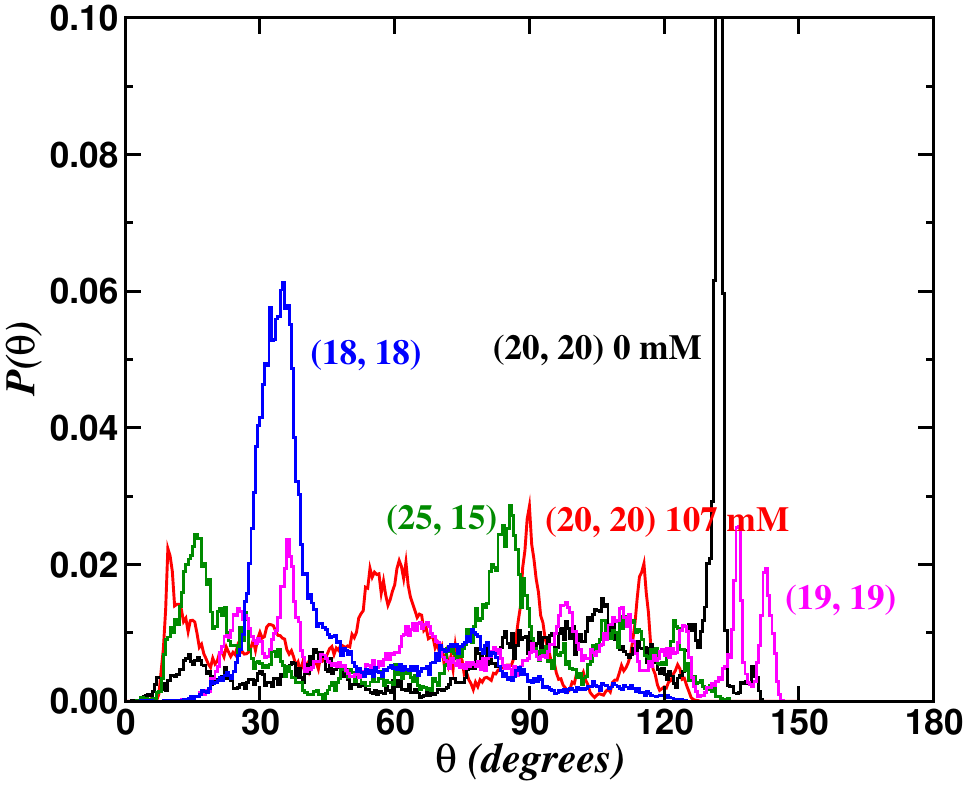}
        \label{rotation}
        }
\caption{Structural deformation: (a) RMSd and 
(b) WC H-bonds of siRNA as a function of time.
These structural parameters are fluctuating before
complete translocation and reaches stable constant
value after the complete translocation of siRNA. This
also indicates that siRNA-CNT hybrid is stable after
translocation. (c) Probability of rotation angle, $\theta$
of siRNA.}
\label{deformations}
\end{figure*}

In order to translocate, siRNA has to cross 
the free energy barrier arising due to loss of 
entropy of siRNA. In this process, many structural 
deformations occurs in siRNA. We have calculated 
several quantities that quantify the degree of structural 
deformations in siRNA. In Figure \ref{rmsd}, we 
we plot the root mean square deviation (RMSd) as 
a function of time as siRNA translocates in the 
interior of CNT for various CNT diameters as 
well as for different salt concentrations. The 
RMSd was calculated with respect to the initial 
minimized structure of siRNA. The RMSd
of the siRNA which is inside of a thinner CNT
is larger than that of the siRNA which is inside
a fatter CNT. This difference is due to the 
large structural deformation of siRNA inside 
CNT of smaller diameter. The average RMSd of 
siRNA is 3.0 \AA, 5.4 \AA~ and 8.2 \AA~ inside 
(20, 20), (19, 19) and (18, 18) CNT, respectively.
Also RMSd for siRNA in (20, 20) CNT is higher 
at 107 mM salt concentration compared to 0 mM 
concentration. Among all the cases, siRNA in 
(20, 20) CNT at 0 mM has least RMSd after 
translocation with small fluctuations.
In the delivery application, the translocation 
method is more appropriate and diameter of the 
CNT can be appropriately chosen to have minimum 
deformation of siRNA structure.\\

To translocate inside smaller diameter CNTs,
siRNA has to get unzipped. To quantify the 
amount of unzipping of siRNA inside CNT, we 
have calculated the number of WC H-bonds 
based on geometry criteria. WC H-bonds play 
a crucial role in maintaining the double 
stranded form of nucleic acids 
{and their function} \cite{voet,santoshjpcm,santoshbj}. For WC 
H-bond calculation, we have used geometry based 
criteria wherein for H-bond, donor-acceptor (D-A)
distance $\le$ 2.7 \AA~ and angle $\angle{\text{DHA}} \ge 130^{\circ}$.
We have plotted the WC H-bonds of siRNA in 
various CNTs in Figure \ref{wc}. There are 
only three broken H-bonds out of total 48 
possible H-bonds for siRNA in the case of 
(20, 20) CNT at 0 and 107 mM {salt concentrations} and dsDNA in 
the case of (20, 20) CNT {at 0 mM salt concentration}. As $d$ is decreasing, 
more H-bonds are broken in siRNA due to stronger
vdW interaction between siRNA and CNT. 
Interestingly, large number of H-bonds of 
siRNA are broken inside (25, 15) CNT.
The aromatic rings of (25, 15) CNT are oriented
differently with respect to CNT axis $\hat{n}$.
Hence nucleobases try to orient along the aromatic 
ring direction due to vdW interaction resulting 
in large number of {broken} H-bonds. 
The larger number of broken H-bonds for 
smaller diameter is also reflected in the 
larger RMSd of siRNA {in case of} smaller diameter 
as shown in Figure \ref{rmsd}.\\

Another very interesting observation is the 
rotation of siRNA to optimize the nucleobase
orientation with respect to the aromatic rings
of CNT while translocating inside 
CNT. We have calculated the distribution of 
rotation {angle}, $P(\theta)$ of siRNA where $\theta$
is the average rotation angle of
siRNA as it translocates inside nanotube with
respect to the structure outside nanotube.
The rotation angle $\theta$ is calculated using
\begin{equation}
\theta(t) = \left<{\cos}^{\text{-1}} \left({\bf{p}}_i\left(t\right)\cdot{\bf{p}}_i\left(0\right)\right)\right>_i
\end{equation} 
where ${\bf{p}}_i(t)$ is the vector joining phosphate-phosphate 
atoms of $i^{th}$ base-pair for $i$ = 4, 5 \dots 17 at time $t$. 
The angular brackets $\left<\dots\right>_i$ denote average
over $i$ base-pairs. Since sticky-ends and
base-pairs near to both the ends of siRNA have 
large fluctuations compared to the middle part of siRNA,
they are omitted for the calculation of $\theta$.
$P(\theta)$ for siRNA while translocation 
inside various CNTs is plotted in Figure \ref{rotation}.
We find that siRNA has to rotate in order to
translocate inside CNT.
Several mechanisms of DNA packaging propose
the rotation of DNA during translocation 
\cite{hendrix,nummela,hugel}.
By rotating, siRNA get reoriented with respect 
to the CNT inner surface which helps in
overcoming local free energy barriers
and helps in the translocation. Rotation is 
large for siRNA inside (20, 20) CNT at 0 mM 
salt concentration and is least for siRNA inside 
(18, 18) CNT with $P(\theta)$ having very 
strong peak at 132$^{\circ}$ and 35$^{\circ}$, 
respectively. {In the case of
other CNTs}, $P(\theta)$ has more than one peak with 
comparable magnitudes indicating
that those rotations are equally likely to be 
observed. Hence, the local free energy barriers 
are escaped by rotation of siRNA in order to 
translocate inside CNTs.
 
\section{Conclusion}
\label{section_conclusions}
To conclude, all atom molecular dynamics simulations 
with explicit solvent are used to investigate the 
translocation {and encapsulation} of siRNA inside CNTs of various 
diameters, chiralities at various salt
concentrations. After the translocation, the 
siRNA stays inside without any further movement.
Free energy calculation using Umbrella sampling 
shows that siRNA gains in free energy while 
translocating inside CNT and has to overcome 
large free energy barrier to exit from the interior of the nanotube.
There is no translocation of siRNA
inside (17, 17) CNT which has a diameter of 22.73 \AA.
We find that the diameter of the (18, 18) CNT is the critical 
diameter for the translocation. Interestingly, 
siRNA translocates inside (20, 20) CNT but 
dsDNA cannot. This difference is 
due to (i) more interaction strength of uridine with
CNT inner surface than that of thymidine \cite{santoshjcp,santoshgraphene}
 and (ii) strong A-T base-pairing 
energy \cite{sponer2004,santoshjcp,santoshgraphene,huang2011} compared to A-U base-pairing
energy. The translocation time $\tau$ is decreasing with 
increasing diameter of the CNT with a critical
diameter of 24 \AA. Inside CNTs of smaller diameters,
vdW repulsion is larger which causes more structural
deformations in siRNA. Free energy barrier for 
exit as well as translocation time $\tau$ strongly 
depends on the chirality of the nanotube. 
The aromatic ring orientation in (25, 15) 
CNT also induces large structural deformations
in siRNA and makes the siRNA and CNT interaction 
stronger than the (20, 20) CNT having same diameter.
This makes the translocation in (25, 15) CNT 
faster compared to (20, 20) CNT. Among all the 
systems studied, siRNA has least
deformations when translocated inside (20, 20) CNT.
The stable siRNA-CNT hybrid can be used to deliver 
siRNA in RNAi applications. The kinetics 
and other thermodynamic analysis presented 
in this work allow us to get a microscopic 
understanding of the translocation process.
By means of rotation, siRNA escapes from the 
local free energy barriers and translocate 
inside nanotubes. 
Modeling CNT with partial atomic charges on carbon
atoms, functionalizing siRNA with polar nanoparticles
for the efficient drug delivery systems are among
exciting future perspectives in this emerging area.

\section{Acknowledgements}
We thank 
department of biotechnology (DBT), 
India for the financial support.

\begin{table}[b]
\caption{Summary of the simulation setup}
        \begin{tabular}{*{8}{|c}|}
       \hline
        (m, n) CNT    & siRNA& Box Dimensions $(\text{\AA}^{3})$ & Na$^+$ & Cl$^-$  & $c~(mM)$         & WAT & Total Atoms \\
       \hline
        (20, 20) 2880 & 1396 & $74\times74\times320$ & 44  & 0   & 0 [NaCl]       & 47619 & 147177   \\
        (20, 20) 2880 & 1396 & $74\times74\times320$ & 157 & 113 & 107 [NaCl]     & 47395 & 146731   \\
        (19, 19) 2736 & 1396 & $73\times73\times318$ & 44  & 0   & 0 [NaCl]       & 45897 & 141867   \\
        (18, 18) 2592 & 1396 & $72\times72\times319$ & 44  & 0   & 0 [NaCl]       & 44393 & 137211   \\
        (17, 17) 2448 & 1396 & $70\times70\times320$ & 44  & 0   & 0 [NaCl]       & 42690 & 131958   \\
        (25, 15) 2940 & 1396 & $74\times74\times318$ & 44  & 0   & 0 [NaCl]       & 47358 & 146454   \\
        (20, 20) 2880 & 1398 & $74\times74\times321$ & 42  & 0   & 0 dsDNA [NaCl] & 47849 & 147867   \\
       \hline
        \end{tabular}
\label{table1}
\end{table}


\begin{thebibliography}{99}
\bibitem{liu1}
Z.~Liu, M.~Winters, M.~Holodniy, and H.~Dai.
\newblock {\em {Angew. Chem. Int. Ed.}}, {46}({12}):{2023--2027}, {(2007)}.

\bibitem{liu3}
Z.~Liu, S.~Tabakman, K.~Welsher, and H.~Dai.
\newblock {\em {Nano Research}}, {2}({2}):{85--120}, {(2009)}.

\bibitem{kam}
N.~Kam, Z.~Liu, and H.~Dai.
\newblock {\em {Angew. Chem. Int. Ed.}}, {45}({4}):{577--581}, {(2006)}.

\bibitem{zhang}
Z.~Zhang, X.~Yang, Y.~Zhang, B.~Zeng, Z.~Wang, T.~Zhu, R.~B.~S. Roden, Y.~Chen,
  and R.~Yang.
\newblock {\em {Clin. Cancer. Res.}}, {12}({16}):{4933--4939}, {(2006)}.

\bibitem{lu2004}
Q.~Lu, J.~Moore, G.~Huang, A.~Mount, A.~Rao, L.~Larcom, and P.~Ke.
\newblock {\em {Nano Lett.}}, {4}({12}):{2473--2477}, {(2004)}.

\bibitem{santoshjcp}
M.~Santosh, S.~Panigrahi, D.~Bhattacharyya, A.~K. Sood, and P.~K. Maiti.
\newblock {\em {J. Chem. Phys.}}, {136}({24}):{065106}, {(2012)}.

\bibitem{santoshgraphene}
S.~Mogurampelly, S.~Panigrahi, D.~Bhattacharyya, A.~K. Sood, and P.~K.~Maiti.
\newblock {\em J. Chem. Phys.}, 137(5):054903, (2012).

\bibitem{hemant_softmatter}
H.~Kumar, Y.~Lansac, M.~A. Glaser, and P.~K. Maiti.
\newblock {\em {Soft Matter}}, {7}({13}):{5898--5907}, {(2011)}.

\bibitem{kasianowicz1996}
J.~Kasianowicz, E.~Brandin, D.~Branton, and D.~Deamer.
\newblock {\em {Proc. Natl. Acad. Sci. USA.}}, {93}({24}):{13770--13773}, 
  {(1996)}.

\bibitem{henrickson}
S.~Henrickson, M.~Misakian, B.~Robertson, and J.~Kasianowicz.
\newblock {\em {Phys. Rev. Lett.}}, {85}({14}):{3057--3060}, {(2000)}.

\bibitem{lubenskinelson1999}
D.~Lubensky and D.~Nelson.
\newblock {\em {Biophys. J.}}, {77}({4}):{1824--1838}, {(1999)}.

\bibitem{meller2001}
A.~Meller, L.~Nivon, and D.~Branton.
\newblock {\em {Phys. Rev. Lett.}}, {86}({15}):{3435--3438}, {(2001)}.

\bibitem{sung1996}
W.~Sung and P.~Park.
\newblock {\em {Phys. Rev. Lett.}}, {77}({4}):{783--786}, {(1996)}.

\bibitem{muthukumar1999}
M.~Muthukumar.
\newblock {\em {J. Chem. Phys.}}, {111}({22}):{10371--10374}, {(1999)}.

\bibitem{muthukumar2001}
M.~Muthukumar.
\newblock {\em {Phys. Rev. Lett.}}, {86}({14}):{3188--3191}, {(2001)}.

\bibitem{ceesdekker2007}
C.~Dekker.
\newblock {\em {Nature Nanotechnol.}}, {2}({4}):{209--215}, {(2007)}.

\bibitem{branton2008}
D.~Branton, D.~W. Deamer, A.~Marziali, H.~Bayley, S.~A. Benner, T.~Butler,
  M.~Di~Ventra, S.~Garaj, A.~Hibbs, X.~Huang, S.~B. Jovanovich, P.~S. Krstic,
  S.~Lindsay, X.~S. Ling, C.~H. Mastrangelo, A.~Meller, J.~S. Oliver, Y.~V.
  Pershin, J.~M. Ramsey, R.~Riehn, G.~V. Soni, V.~Tabard-Cossa, M.~Wanunu,
  M.~Wiggin, and J.~A. Schloss.
\newblock {\em {Nature Biotechnol.}}, {26}({10}):{1146--1153}, {(2008)}.

\bibitem{ceesdekker2006}
R.~Smeets, U.~Keyser, D.~Krapf, M.~Wu, N.~Dekker, and C.~Dekker.
\newblock {\em {Nano Lett.}}, {6}({1}):{89--95}, {(2006)}.

\bibitem{ceesdekker2009}
S.~van Dorp, U.~F. Keyser, N.~H. Dekker, C.~Dekker, and S.~G. Lemay.
\newblock {\em {Nature Phys.}}, {5}({5}):{347--351}, {(2009)}.

\bibitem{gao2003}
H.~Gao, Y.~Kong, D.~Cui, and C.~Ozkan.
\newblock {\em {Nano Lett.}}, {3}({4}):{471--473}, {(2003)}.

\bibitem{fan2005nl}
R.~Fan, R.~Karnik, M.~Yue, D.~Li, A.~Majumdar, and P.~Yang.
\newblock {\em {Nano Lett.}}, {5}({9}):{1633--1637}, {(2005)}.

\bibitem{gao2007}
Y.~Xie, Y.~Kong, A.~K. Soh, and H.~Gao.
\newblock {\em {J. Chem. Phys.}}, {127}({22}):{225101}, {(2007)}.

\bibitem{gao2008}
Q.~X. Pei, C.~G. Lim, Y.~Cheng, and H.~Gao.
\newblock {\em {J. Chem. Phys.}}, {129}({12}):{125101}, {(2008)}.

\bibitem{lim2008}
M.~C.~G. Lim, Q.~Pei, and Z.~W. Zhong.
\newblock {\em {Physica A}}, {387}({13}):{3111--3120}, {(2008)}.

\bibitem{liuscience2010}
H.~Liu, J.~He, J.~Tang, H.~Liu, P.~Pang, D.~Cao, P.~Krstic, S.~Joseph,
  S.~Lindsay, and C.~Nuckolls.
\newblock {\em {Science}}, {327}({5961}):{64--67}, {(2010)}.

\bibitem{garaj}
S.~Garaj, W.~Hubbard, A.~Reina, J.~Kong, D.~Branton, and J.~A. Golovchenko.
\newblock {\em {Nature}}, {467}({7312}):{190--193}, {(2010)}.

\bibitem{schneider}
G.~F. Schneider, S.~W. Kowalczyk, V.~E. Calado, G.~Pandraud, H.~W. Zandbergen,
  L.~M.~K. Vandersypen, and C.~Dekker.
\newblock {\em {Nano Lett.}}, {10}({8}):{3163--3167}, {(2010)}.

\bibitem{christopher}
C.~A. Merchant, K.~Healy, M.~Wanunu, V.~Ray, N.~Peterman, J.~Bartel, M.~D.
  Fischbein, K.~Venta, Z.~Luo, A.~T.~C. Johnson, and M.~Drndic.
\newblock {\em {Nano Lett.}}, {10}({8}):{2915--2921}, {(2010)}.

\bibitem{venkatesan2011}
B.~M. Venkatesan and R.~Bashir.
\newblock {\em {Nature Nanotechnol.}}, {6}({10}):{615--624}, {(2011)}.

\bibitem{torrie}
G.~Torrie and J.~Valleau.
\newblock {\em {J. Comput. Phys.}}, {23}({2}):{187--199}, {(1977)}.

\bibitem{roux}
B.~Roux.
\newblock {\em {Comput. Phys. Commun.}}, {91}({1-3}):{275--282}, {(1995)}.

\bibitem{ums}
{{D.~Frenkel, and B.~Smit.}
\newblock {{2} ed.; {Academic Press}}, {(2001)}.}

\bibitem{duan}
Y.~Duan, C.~Wu, S.~Chowdhury, M.~Lee, G.~Xiong, W.~Zhang, R.~Yang, P.~Cieplak,
  R.~Luo, T.~Lee, J.~Caldwell, J.~Wang, and P.~Kollman.
\newblock {\em {J. Comput. Chem.}}, {24}({16}):{1999--2012}, {(2003)}.

\bibitem{maiti2004}
P.~K. Maiti, T.~A. Pascal, N.~Vaidehi, and W.~A. Goddard, III.
\newblock {\em {Nuc. Acids Res.}}, {32}({20}):{6047--6056}, {(2004)}.

\bibitem{maiti2006bj}
P.~K. Maiti, T.~A. Pascal, N.~Vaidehi, J.~Heo, and W.~A. Goddard, III.
\newblock {\em {Biophys. J.}}, {90}({5}):{1463--1479}, {(2006)}.

\bibitem{darden}
T.~Darden, D.~York, and L.~Pedersen.
\newblock {\em {J. Chem. Phys.}}, {98}({12}):{10089--10092}, {(1993)}.

\bibitem{yildirim}
I.~Yildirim, H.~A. Stern, S.~D. Kennedy, J.~D. Tubbs, and D.~H. Turner.
\newblock {\em J. Chem. Theory Comput.}, 6(5):1520--1531, (2010).

\bibitem{lin2003}
S.-T.~Lin, M.~Blanco, and W.~A.~Goddard, III.
\newblock {\em {J. Chem. Phys.}}, {119}({22}):{11792--11805}, {(2003)}.

\bibitem{lin2010}
S.-T. Lin, P.~K. Maiti, and W.~A. Goddard, III.
\newblock {\em {J. Phys. Chem. B}}, {114}({24}):{8191--8198}, {(2010)}.

\bibitem{pascal2011}
{T.~A. Pascal, W.~A. Goddard, III, and Y.~Jung.
\newblock {\em {Proc. Natl. Acad. Sci. USA.}}, {108}({29}):{11794--11798}, {(2011)}.}

\bibitem{still1990}
{W.~C. Still, A.~Tempczyk, R.~C. Hawley, and T. Hendrickson.
\newblock {\em {J. Am. Chem. Soc.}}, {112}({16}):{6127--6129}, {(1990)}.}

\bibitem{srinivasan1999}
{J.~Srinivasan, M.~W. Trevathan, P.~ Beroza, and D.~A. Case.
\newblock {\em {Theor. Chem. Acc.}}, {101}({6}):{426--434}, {(1999)}.}

\bibitem{maitinl}
{P.~K. Maiti, and B. Bagchi.}
\newblock {\em {Nano Lett.}}, {6}:{2478--2485}, {(2006)}.

\bibitem{hemant}
H.~Kumar, B.~Mukherjee, S.-T. Lin, C.~Dasgupta, A.~K. Sood, and P.~K. Maiti.
\newblock {\em {J. Chem. Phys.}}, {134}({12}):{124105}, {(2011)}.

\bibitem{nandy2011}
{B.~Nandy, and P.~K.~Maiti.}
\newblock {\em {J. Phys. Chem. B}}, {115}:{217--230}, {(2011)}.

\bibitem{supplementary}
See Supplementary Material Document No. --- for the details of 
{non-equilibrium versus equilibrium free energy methods and for more snapshots}.

\bibitem{sponer2004}
{Sponer, J.; Jurecka, P.; Hobza, P.}
\newblock {\em {J. Am. Chem. Soc.}}, {126}:{10142--10151}, {(2004)}.

\bibitem{huang2011}
Y.~Huang, X.~Weng, and I.~M. Russu.
\newblock {\em {Biochemistry}}, {50}({11}):{1857--1863}, {(2011)}.

\bibitem{ito2003}
T.~Ito, L.~Sun, and R.~Crooks.
\newblock {\em {Chem. Commun.}}, ({13}):{1482--1483}, {(2003)}.

\bibitem{santoshjbs}
B.~Nandy, M.~Santosh, and P.~K. Maiti.
\newblock {\em {J. Biosci.}}, {37}({3}):{457--474}, {(2012)}.

\bibitem{voet}
D.~Voet and J.~G. Voet.
\newblock John Wiley \& Sons. Inc.,, 3 edition, (2005).

\bibitem{santoshjpcm}
M.~Santosh and P.~K. Maiti.
\newblock {\em {J. Phys.: Condens. Matter}}, {21}({3}):{034113}, {(2009)}.

\bibitem{santoshbj}
M.~Santosh and P.~K. Maiti.
\newblock {\em {Biophys. J.}}, {101}({6}):{1393--1402}, {(2011)}.

\bibitem{hendrix}
R.~Hendrix.
\newblock {\em {Proc. Natl. Acad. Sci. USA.}}, {75}({10}):{4779--4783}, {(1978)}.

\bibitem{nummela}
J.~Nummela and I.~Andricioaei.
\newblock {\em {Biophys. J.}}, {96}({4}):{L29--L31}, {(2009)}.

\bibitem{hugel}
T.~Hugel, J.~Michaelis, C.~L. Hetherington, P.~J. Jardine, S.~Grimes, J.~M.
  Walter, W.~Faik, D.~L. Anderson, and C.~Bustamante.
\newblock {\em {PLoS Biol}}, {5}({3}):{558--567}, {(2007)}.

\bibitem{humphrey}
W.~Humphrey, A.~Dalke, and K.~Schulten.
\newblock {\em {J. Mol. Graph.}}, {14}({1}):{33--\&}, {(1996)}.

\end{thebibliography}
\end{document}


\author{Santosh Mogurampelly}\thanks{To whom correspondence should be addressed}
\email{santosh@physics.iisc.ernet.in}
\author{Prabal K.~Maiti}
\email{maiti@physics.iisc.ernet.in}
\affiliation{
Centre for Condensed Matter Theory, Department of Physics, Indian Institute of Science,
Bangalore 560012, India}
\title{Translocation and encapsulation of siRNA inside carbon nanotubes}
\maketitle
\pagebreak

\section{Non-equilibrium versus equilibrium free energy methods}
\begin{figure*}
        \centering
        \subfigure[]
        {
        \begin{overpic}[height=60mm]{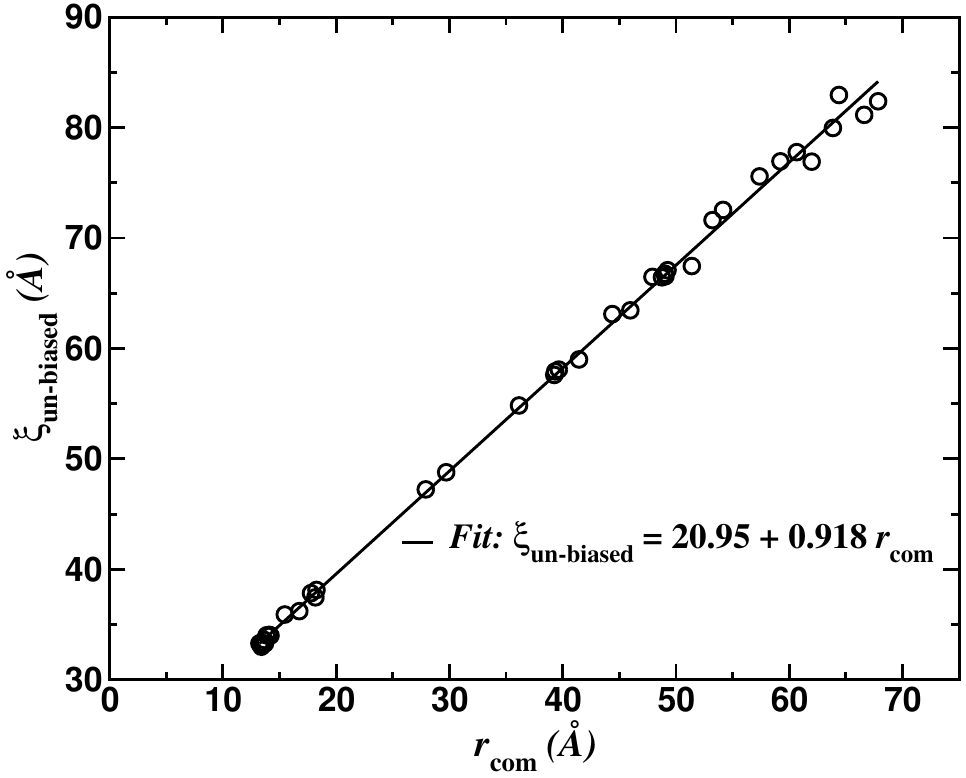}
        \put(48,54){$r_{\text{com}}$}
        \put(15,57){\includegraphics[height=15mm]{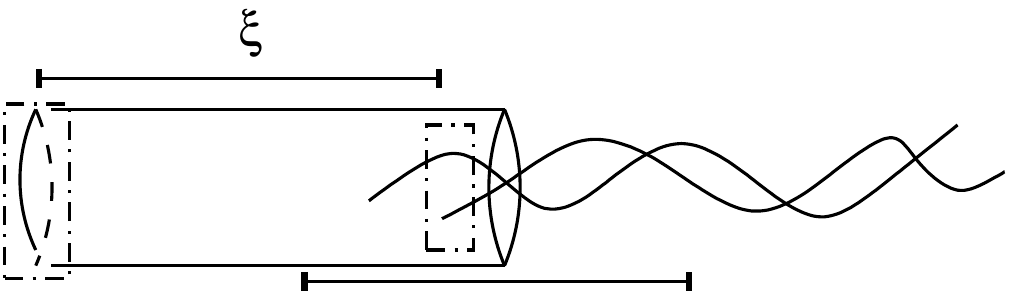}}
        \end{overpic}
        \label{zeta_com}
	}
        \subfigure[]
        {
        \begin{overpic}[height=6cm]{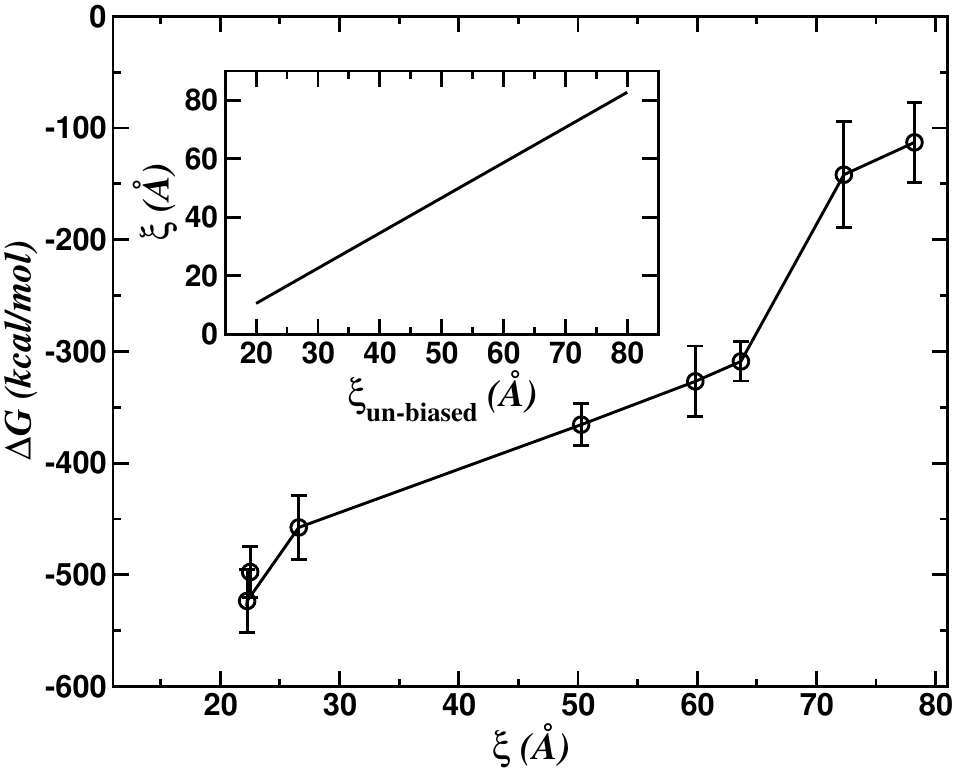}
        \put(71,68){\includegraphics[height=0.44cm]{0ns_horizontal}}
        \put(56,24){\includegraphics[height=0.55cm]{20ns_horizontal}}
        \put(29,13){\includegraphics[height=0.55cm]{35ns_horizontal}}
        \end{overpic}
        \label{free_energy_zeta}
        }
\caption{(a) Dependency
of $\xi_{\text{un-biased}}$ on $r_{\text{com}}$ from 
which $\xi$ can be calculated as $\xi = 1.2
\xi_{\text{un-biased}} -13.456$ as shown in inset 
of Figure \ref{free_energy_zeta}; where $\xi$ is the distance 
between the center of masses of `far end' of CNT and 
first two Watson-Crick (WC) H-bonded base-pairs of 
siRNA near to the CNT as shown in inset cartoon, 
$\xi_{\text{un-biased}}$ is the corresponding parameter 
in un-biased simulations. (b) $\Delta{G}$ as a 
function of reaction coordinate $\xi$ as defined in 
the Umbrella sampling for comparison.}
\label{2pt-mmpbsa}
\end{figure*}

\begin{figure*}
        \centering
        \subfigure[]
        {
        \includegraphics[height=45mm]{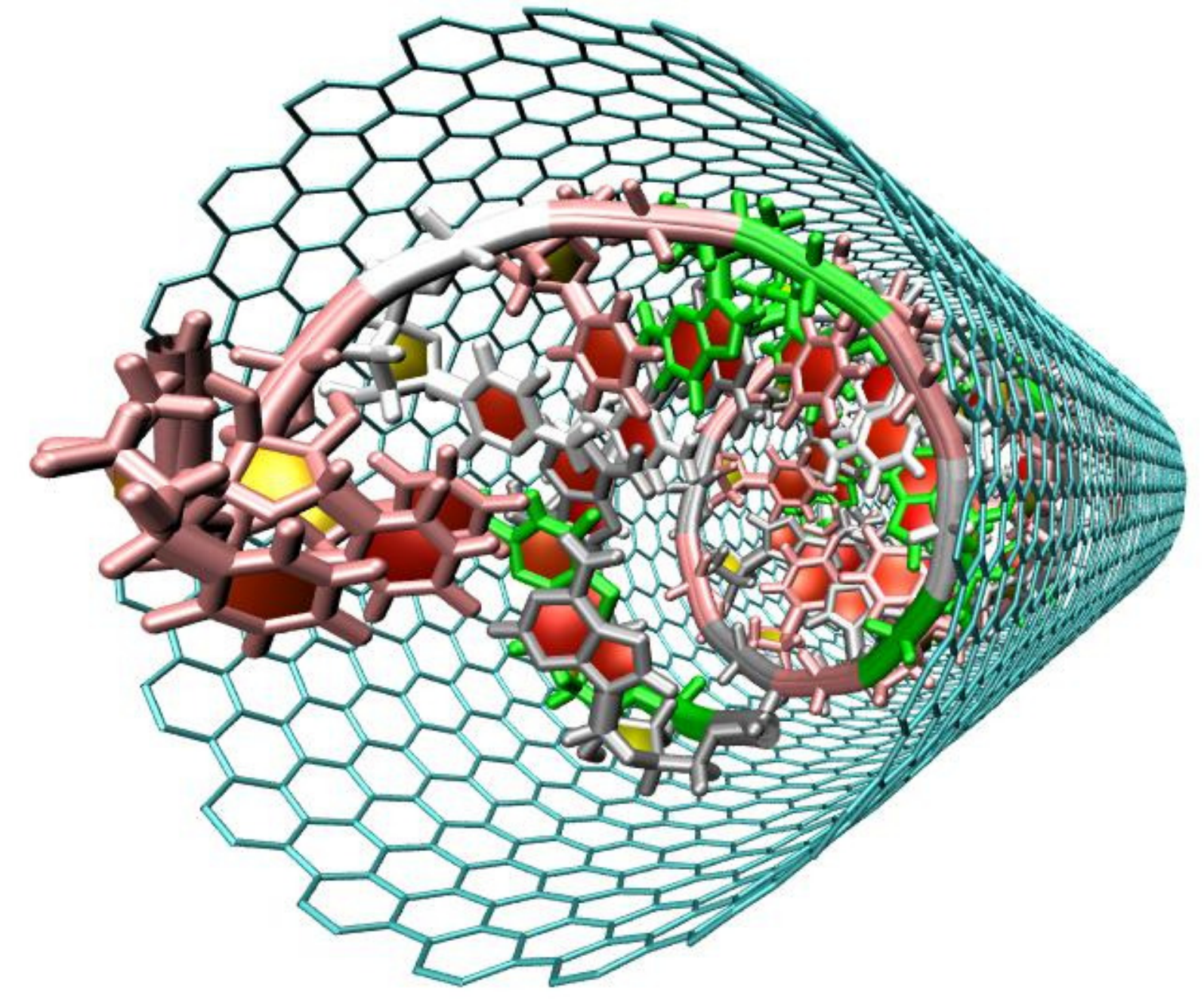}
        \label{40ns_view3}
        }
        \subfigure[]
        {
        \includegraphics[height=45mm]{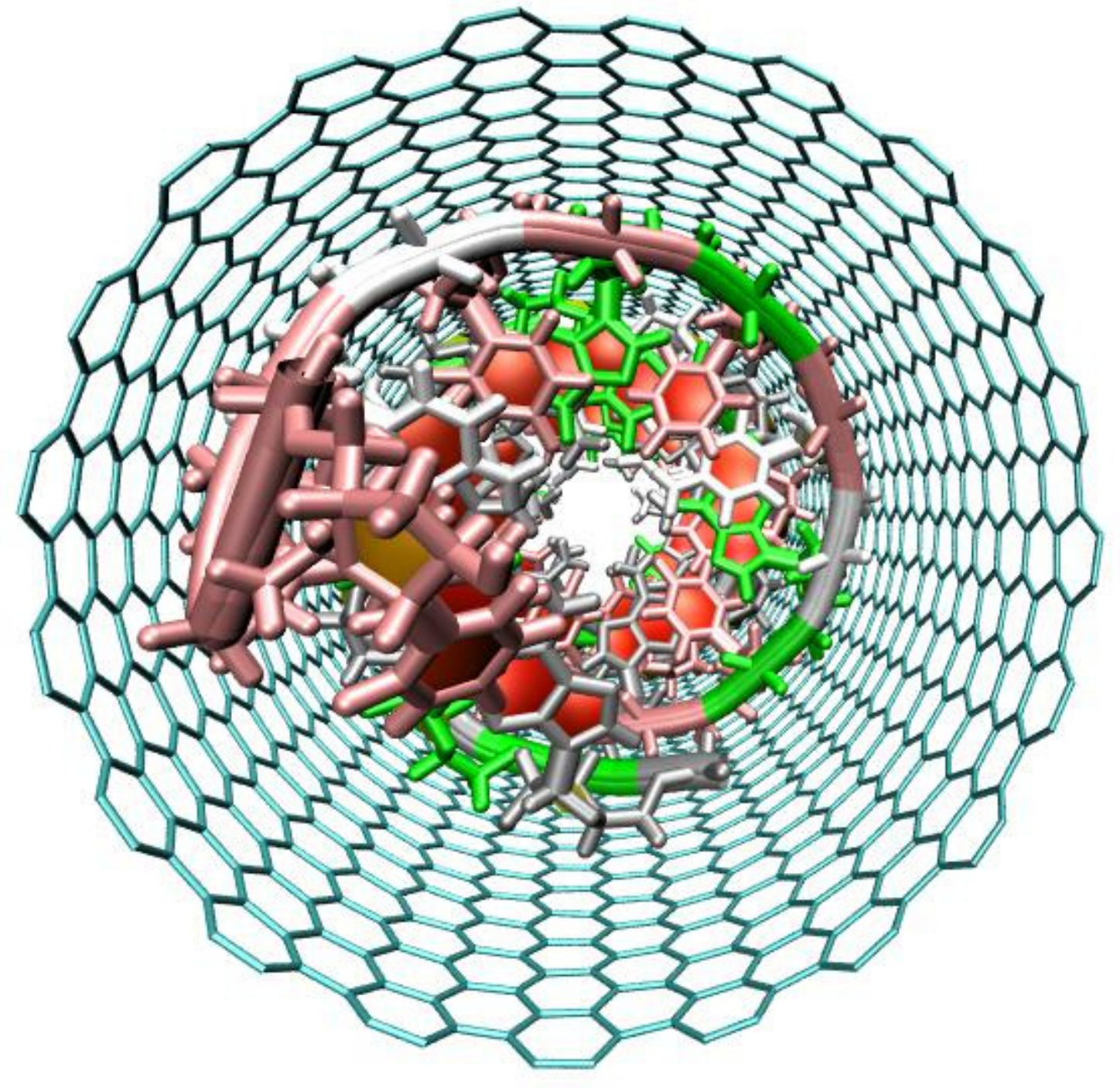}
        \label{40ns_view2}
        }
        \subfigure[]
        {
        \includegraphics[height=45mm]{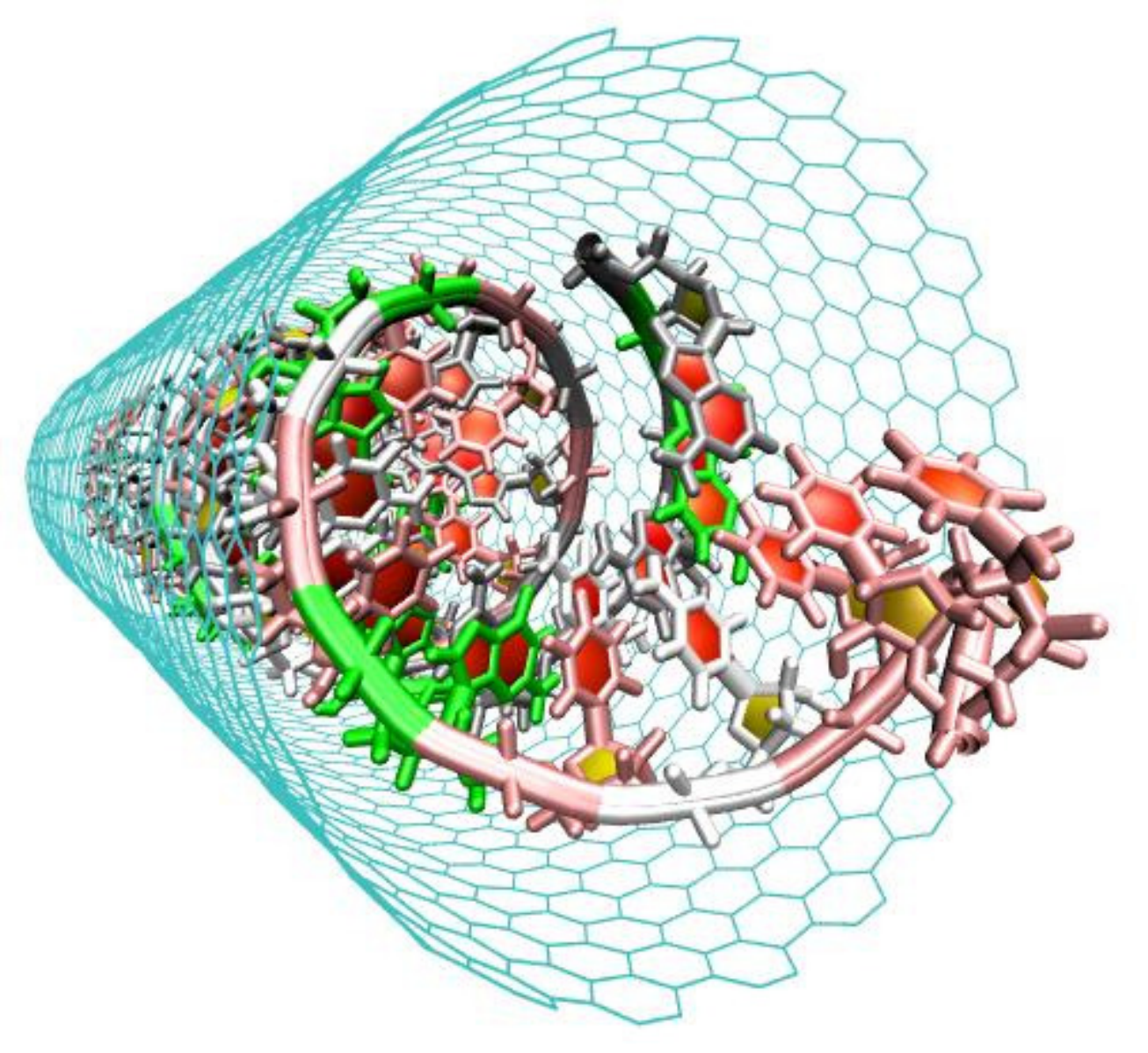}
        \label{40ns_view1}
        }\\
        \subfigure[]
        {
        \includegraphics[height=35mm]{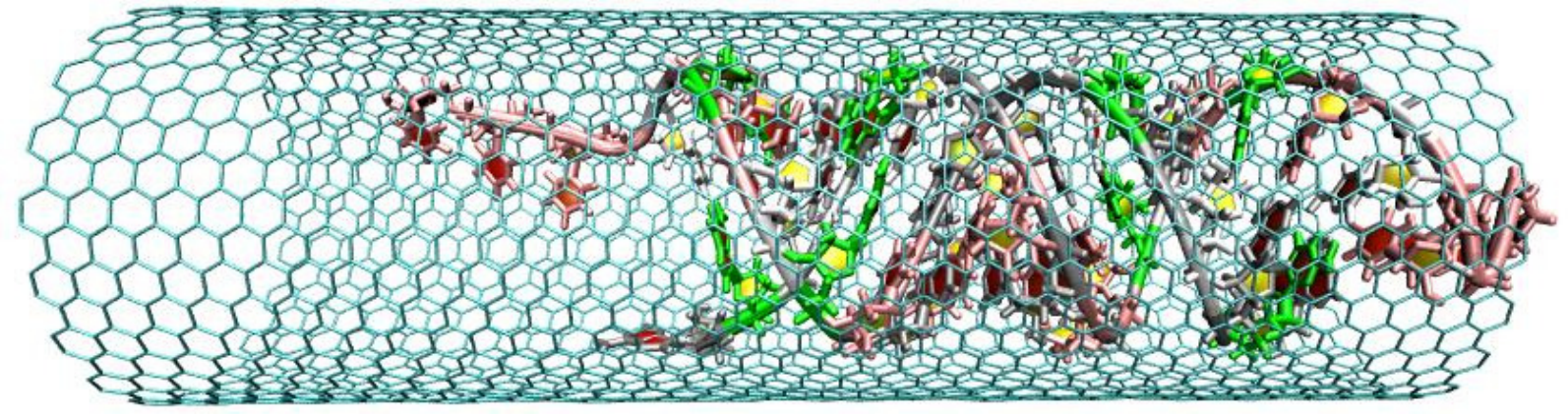}
        \label{40ns_view4}
        }\\
        \caption{Snapshots of siRNA after complete 
translocation inside (20, 20) CNT. These snapshots 
were rendered at 39 ns in perspective display mode.}
        \label{snapshots2}
\end{figure*}

Transformation of binding energy from a function of 
simulation time $t$ to a function of $r_{\text{com}}$, 
center of mass distance between CNT and siRNA is 
carried out (See the schematic in Figure \ref{zeta_com} for 
the pictorial definition of $r_{\text{com}}$). The 
transformation is done simply by 
monitoring $r_{\text{com}}(t)$. For the Umbrella 
sampling calculations, a new distance parameter $\xi$, called 
reaction coordinate is introduced. For simplicity in implementation
without loss of generality and understanding, we calculate
$\xi$ as the distance between the center of masses of `far end' 
of CNT and first two Watson-Crick (WC) H-bonded base-pairs 
of siRNA near to the CNT (See schematic in Figure \ref{zeta_com}). 
For the notations to follow, we use the term `un-biased' 
for the MD simulation that are performed without any 
external biased forces and `biased' for the MD simulations 
with external bias as the case of Umbrella sampling. In 
the un-biased MD simulations, the corresponding parameter 
for $\xi$ is $\xi_{\text{un-biased}}$.
We plot $\xi_{\text{un-biased}}$ as a function of 
$r_{\text{com}}(t)$ in Figure \ref{zeta_com}. 
Due to the effect of Umbrella biasing potential there
could be minor structural deformations in siRNA leading to 
different values of $\xi$ and $\xi_{\text{un-biased}}$ 
for same value of $r_{\text{com}}$ in both the biased and 
un-biased simulations. Using the $\xi$ values from 
Umbrella sampling, we plot $\xi$ as a function of 
$\xi_{\text{un-biased}}$ in the inset of 
Figure \ref{free_energy_zeta}. By assuming (fitting) $\xi$ 
to depend linearly on $\xi_{\text{un-biased}}$; 
{\it{i.e.,}} $\xi = m\xi_{\text{un-biased}} + c$, 
we arrive at $m$ = 1.2022 and $c$ = -13.456.
This calibration procedure helps us to calculate 
$\Delta{G}$ as a function of $\xi$ as shown 
in Figure \ref{free_energy_zeta} which is more 
relevant to compare with Umbrella sampling results.
As can be seen in Figure \ref{free_energy_zeta}, $\Delta{G}$
seems to have a global minima at $\xi$ = 22 \AA~
where siRNA is inside (20, 20) CNT as the inset 
snapshot indicates. At $\xi$ = 78 \AA~ or more, 
siRNA is away from CNT which is at peak of 
unstable configuration. Intermediate 
values of $\xi$ are also shown with snapshots. 
The limitation in calculating $\Delta{G}$ as a function
of $\xi$ described above is that the system is not
well sampled at each $\xi$ to have an ensemble averages
since the translocation process is spontaneous.
To improve the sampling at each $\xi$ to great level 
and obtain $\Delta{G}$ to great accuracy, we employ 
Umbrella sampling method as described in the previous section.
Snapshots of siRNA in (20, 20) CNT are shown in 
Figure \ref{snapshots2} at 40 ns in perspective 
display mode.
\clearpage

\clearpage
